\documentstyle[12pt,aaspp4]{article}

\lefthead{Soo-Chang Rey et al.}
\righthead{Stellar Populations and Age-Metallicity relation of the $\omega$ Centauri}

\begin{document}
\title{CCD PHOTOMETRY OF THE GLOBULAR CLUSTER\\
     $\omega$ CENTAURI.\\
       II. STELLAR POPULATIONS AND AGE-METALLICITY RELATION}

\author{Soo-Chang Rey\altaffilmark{1,} \altaffilmark{2}, Young-Wook Lee, Chang H. Ree, Jong-Myung Joo, 
and Young-Jong Sohn}
\affil{Center for Space Astrophysics \& Department of Astronomy,\\
    Yonsei University, Shinchon 134, Seoul 120-749, Korea\\
        Electronic mail : (screy, ywlee, chr, jmjoo, sohnyj)@csa.yonsei.ac.kr}

\and

\author{Alistair R. Walker}
\affil{Cerro Tololo Inter-American Observatory, NOAO, Casilla 603, La Serena, Chile\\
        Electronic mail : awalker@noao.edu}

\altaffiltext{1}{Current Address: California Institute of Technology, 1200 E. California Blvd, 
MC 405-47, Pasadena, CA 91125; screy@srl.caltech.edu.}
\altaffiltext{2}{Visiting Astronomer, Cerro Tololo Inter-American Observatory,
National Optical Astronomy Observatories, which is operated by the Association
of Universities for Research in Astronomy, Inc. (AURA) under cooperative agreement
with the National Science Foundation.}

\begin{abstract}

We present wide-field and high-precision $BV$ and $Ca$ \& Str\"omgren $by$ photometry 
of $\omega$~Centauri, which represents one of the most extensive photometric surveys 
to date for this cluster. The member stars of $\omega$~Cen are well discriminated from 
foreground Galactic field stars in the $hk$ [=$(Ca-b)-(b-y)$] vs. $b-y$ diagram. 
The resulting ``cleaned" color-magnitude diagram (CMD) has allowed us to obtain an
accurate distribution of the red horizontal branch (HB) and the asymptotic giant branch stars. 
We confirm the presence of several red giant branches (RGBs) with the most metal-rich
sequence well separated from other bluer metal-poor ones. Our population models suggest 
that four populations with different metallicities can reproduce the observed 
nature of the RGB. The HB distribution is also found to be consistent with the multiple 
stellar populations of the RGB. From our population models, we propose that the most 
metal-rich population is about 4 Gyr younger than the dominant metal-poor population, 
indicating that $\omega$ Cen was enriched over this timescale. We identify, 
for the first time, a continuous and slanting RGB bump in the CMD of $\omega$ Cen, 
which is due to the metallicity spread amongst the RGB stars.  Our photometry also 
reveals a significant population of blue straggler stars. The discovery of several 
populations and the internal age-metallicity relation of $\omega$ Cen provides good 
evidence that $\omega$~Cen was once part of a more massive system that merged with 
the Milky Way, as the Sagittarius dwarf galaxy is in the process of doing at the present 
time.

\end{abstract}


\section{INTRODUCTION}

The cluster $\omega$ Centauri (NGC 5139) has many unique characteristics amongst Galactic 
globular clusters (GGCs). It is the most luminous and massive cluster in the Galaxy 
(Harris 1996; Meylan et al. 1995), and one of the most flattened clusters (Meylan 1987; 
White \& Shawl 1987) which must have resulted from its significant rotation 
(Meylan \& Mayor 1986; Mayor et al. 1997; Merrit, Meylan, \& Mayor 1997). The most intriguing 
and peculiar feature of $\omega$ Cen is the wide spread of metallicity. The majority of GGCs 
show homogeneity in iron-peak elements, while often exhibiting large variations in the lighter 
elements such as CNO, Na, Mg, and Al (Suntzeff 1993; Kraft 1994). However, unlike other GCs, 
$\omega$ Cen shows star-to-star abundance variations of the iron-peak elements. 

Dickens \& Woolley (1967) first noted the chemical inhomogenity of stars in $\omega$ Cen, 
from the large color width of the red giant branch (RGB) stars in the CMD shown by the 
photographic photometry of Woolley (1966), which was later confirmed by Cannon \& Stobie (1973). 
Freeman \& Rodgers (1975) and Butler, Dickens, \& Epps (1978) also confirmed a diversity of 
chemical composition from the RR Lyrae variables in this cluster 
(see also Gratton, Tornambe, \& Ortolani 1986; Rey et al. 2000). Recently, a series of 
spectroscopic studies for large samples of the RGB stars in $\omega$ Cen verified the range 
in abundances and derived detailed abundance patterns (e.g., Brown \& Wallerstein 1993; 
Vanture, Wallerstein, \& Brown 1994; Norris \& Da Costa 1995; Smith, Cunha, \& Lambert 1995; 
Norris, Freeman, \& Mighell 1996; Suntzeff \& Kraft 1996; Smith et al. 2000; Pancino et al. 2002; 
Cunha et al. 2002; Origlia et al. 2003). It has been reported that the abundance of iron spans 
a range from [Fe/H] $\sim$ -2.0 up to -0.5, along with high abundance of the s-process elements 
over the whole [Fe/H] range.

It also appears that this anomaly of chemical inhomogeneity found in $\omega$ Cen is closely 
linked to the kinematics and spatial distribution of the stars. Norris et al. (1997) found that 
the metal-rich RGB stars are more centrally concentrated and furthermore show evidence for 
different velocity dispersion and rotation properties compared to the dominating population 
of more metal-poor stars.  The metal-poor population rotates, but the metal-rich stars do not.  
From this apparent difference in kinematics between the metal-poor and the metal-rich stars, 
as well as a second peak in the metallicity distribution, they suggested a merger of two clusters 
as a possible origin of $\omega$ Cen. The difference in spatial distribution between different 
metallicity populations is also confirmed by successive studies (Jurcsik 1998; 
Pancino et al. 2000, 2003; Hilker \& Richtler 2000; see also Freeman 1985). Most recently, 
Ferraro, Bellazzini, \& Pancino (2002) suggested that the newly discovered metal-rich RGB stars 
with [Fe/H] $\sim$ -0.6 may have a different proper motion distribution that is not compatible 
with that of the dominant metal-poor population of $\omega$ Cen, while Platais et al. (2003) 
commented that this could arise as a spurious effect in the proper motion determination. 
The close connection of metal abundance to the kinematics and spatial distribution of 
stars in $\omega$ Cen would provide some important constraints for its star formation and 
chemical enrichment histories when combined with wide-field high precision photometric data. 

The first comprehensive and wide-field photographic study for $\omega$ Cen was undertaken 
by Woolley (1966). It remained the only wide-field study of $\omega$ Cen for more than three decades. 
He, and co-workers, obtained a CMD for several thousand stars which extends to about 1 mag fainter 
than the horizontal branch (HB), however the photometric errors for even bright RGB stars were 
somewhat large. These data are too shallow to allow accurate study of the properties of various 
populations in this cluster, despite being spatially complete.  Recent wide field studies were
initiated by Lee et al. (1999), who obtained high-precision homogeneous $BV$ CCD CMDs for 
more than 130,000 stars in the field toward $\omega$~Cen, which represents one of the most 
extensive photometric surveys to date for this cluster (see also Pancino et al. 2000; 
Hilker \& Richtler 2000; Majewski et al. 2000). Lee et al. (1999) discovered multiple RGBs, 
especially noteworthy being a red, metal-rich ([Fe/H] $\sim$ -0.5) sequence well separated from 
other, bluer, metal-poor ones in the CMD. This feature was not evident in previous photometry 
due to larger photometric uncertainties (e.g., Woolley 1966) and usually much smaller sample sizes. 
An independent survey by Pancino et al. (2000) confirmed the reality of this discovery. 
Furthermore, the most metal-rich population in $\omega$~Cen appears to be a few billion years 
younger than the most metal-poor ([Fe/H] $\sim$ -1.8) population in this system 
(Lee et al. 1999; Hughes \& Wallerstein 2000; Hilker \& Richtler 2000).

The multimodal  metallicity distribution function and the apparent age-metallicity relation 
would suggest that the protocluster of $\omega$~Cen was massive enough to undergo some 
self-enrichment (e.g., Suntzeff \& Kraft 1996; Ikuta \& Arimoto 2000) and several early bursts 
of star formation with a variable rate. This suggests that $\omega$~Cen evolved within a 
dwarf galaxy size gas-rich subsystem until it merged and disrupted with our Galaxy a few billion 
years after the formation of its first generation of metal-poor stars, leaving its core as today's 
GC $\omega$~Cen (Freeman 1993; Norris et al. 1997; Lee et al. 1999; Majewski et al. 2000; 
Hughes \& Wallerstein 2000; Hilker \& Richtler 2000). This is consistent with the fact that 
there is similarity in the distinct appearance of multiple stellar populations with an internal 
age-metallicity relation for both $\omega$ Cen and the Sagittarius dwarf galaxy 
(Sarajedini \& Layden 1995; Layden \& Sarajedini 1997, 2000; Bellazzini et al. 1999a, 1999b); 
the latter includes M54, the second most massive GGC as its nucleus and is now in a process of 
tidal disruption by the Milky Way, leading eventually to a complete merger.

In this paper, we discuss our photometry of $\omega$ Cen, which is the same data set used in the 
preliminary study by Lee et al. (1999), analyzing in detail the multiple stellar populations and 
their age-metallicity relations. The use of high precision and wide-field CCD photometry as presented here 
is essential for a better understanding of the characteristics and evolutionary history for 
$\omega$ Cen. We have also found some new properties of the various populations of $\omega$ Cen 
which were not apparent in the  previous photometric studies. In Sec. 2 we describe our observations,
the data reduction, and the photometric calibration. We present the resulting CMDs in Sec. 3 
after subtracting the contamination of the field star sequence using the $hk$ index which is the 
sensitive metallicity indicator of the $Ca$ \& Str\"omgren $by$ photometric system. 
We also present some characteristics of various population sequences, such as the RGB, 
the RGB bump, the blue straggler stars (BSSs), and the HB, in the CMD of $\omega$ Cen. 
In particular, we present the evidence for multiple stellar populations evinced by 
the structure of the RGB. In Sec. 4, we present radial distributions of the RGB, HB, 
and BSS populations. Sec. 5 is devoted to the derivation of internal age-metallicity 
relation of $\omega$ Cen from our population models. In Sec. 6, we discuss the origin of 
$\omega$ Cen and consider the possibility of other massive GCs having a metallicity spread 
and multiple stellar populations like $\omega$ Cen. We summarize our results in Sec. 7.

\section{OBSERVATIONS, DATA REDUCTION, AND PHOTOMETRIC MEASUREMENT}

\subsection {$BV$ Observations}

All the observations were made using the Cerro Tololo Inter-American Observatory (CTIO)
0.9m telescope and Tektronix 2048 No. 3 CCD during six nights of an observing run in
1996 April 5 - 10. At the f/13.5 Cassegrain focus, the image scale of the CCD is 0.396 arcsec/pixel,
which gives a sky coverage of $13.6 \times 13.6$ arcmin$^{2}$. We covered nine regions in a
3 $\times$ 3 grid of $\omega$ Cen through the Johnson $B$ and $V$ filters, and observed 6 to 8
sequences of this grid per night. In total, 40 to 42 frames were taken in each filter and
each field. Our whole program field, centered on the cluster, encompasses approximately a
$40 \times 40$ arcmin$^{2}$ area of $\omega$ Cen, which approximates the area enclosed by 
the half-tidal radius. Exposure times were 180 sec and 120 sec for 
$B$ and $V$, respectively. The seeing was between 1.0 and 1.7 arcsec, and all of the observing 
nights were fully photometric. The positions of all observed fields are given in Table 1 and 
in Figure 1.

The raw CCD frames were processed by zero-level subtraction followed by sky flat-field division
using the IRAF QUADPROC routines to take into account the four amplifiers used. All exposure 
times were sufficiently long so that the center-to-corner shutter timing error was negligible. 
Typically, these procedures produced object frames flattened to better than 0.7\%.
Each of the 40 - 42 CCD frames for each filter and each field was reduced separately using the
DAOPHOT II and ALLSTAR  photometry package (Stetson 1987, 1995). Between 50 and 100 bright and
isolated stars were used to construct the point-spread function (PSF) for each frame. The shape
of the PSF was made to vary quadratically with radial position in the frame. The mean difference 
between the PSF-based magnitude and the total magnitude of selected PSF stars was calculated, and 
this aperture correction was applied to PSF-based instrumental magnitude for every star in each frame. The DAOMATCH/DAOMASTER programs (Stetson 1992) were then used to match stars of all frames for each filter and each field, and to derive their average instrumental magnitudes and rms scatters, 
keeping only those stars that were detected in at least 20 frames for each filter and each field. Finally, the mean instrumental $B$ and $V$ magnitudes were matched to form the $B-V$ color.

On each night, standard stars from the list of Landolt (1992) were observed to derive
coefficients of the transformation equations between the instrumental magnitudes and 
the standard system magnitudes. The 33 standard stars observed cover $B-V$ and $V$ ranges of 
-0.27 to 1.91 and 11.08 to 15.02, respectively. The atmospheric extinction coefficients in each 
color have been determined by the same standard stars at different air masses. The final 
transformation equations were obtained by linear least-square fits. They are 
        $$B-V = 0.905(b-v)_{o} - 0.223,$$
        $$V-v_{o} = 0.021(B-V) - 2.957,$$
where $B-V$ and $V$ are the color indices and visual magnitude in the standard $BV$ system,
$(b-v)_{o}$ and $v_{o}$ refer to instrumental magnitudes corrected for extinction. The calibration 
equations relate observed to standard values with rms residuals of 0.012 mag for both of $V$ 
and $B-V$, as shown in Figure 2.

\subsection {$Ca$ \& Str\"omgren $by$ Observations}
As part of our investigation of the metallicity of RR Lyrae stars in $\omega$ Cen, we have 
obtained $Ca$ \& Str\"omgren $by$ CCD frames with the CTIO 0.9 m telescope and Tektronix 2048 
No. 3 CCD, during three nights of an observing run in 1997 March. These observations cover 
similar fields to those of our $BV$ observations. The details of observations and data reductions for 
these frames are described in Rey et al. (2000). We used these photometric data in order to 
discriminate the foreground Galactic field stars from the $\omega$ Cen member stars using different 
distributions in the $hk$[=$(Ca-b)-(b-y)$] vs. $b-y$ diagram (see Section 3.2).

\subsection {Comparison with Previous Results}
We compared our $BV$ CCD photometry with data taken from other studies for the stars
in common. We made a comparison, shown in Figure 3, with the $V$ and $B-V$ CCD photometry 
of the $\omega$ Cen HST calibration field obtained by Walker (1994). The mean difference 
in the sense our measurements minus Walker's is -0.030 $\pm$ 0.017 and -0.007 $\pm$ 0.016 
in $V$ and $B-V$, respectively, where the uncertainty is the standard deviation of the mean. 
Figure 3 also shows a comparison with photometry of Alcaino \& Liller (1987). The mean differences 
are negligible, 0.007 $\pm$ 0.023 and 0.001 $\pm$ 0.078 in $V$ and $B-V$, respectively, 
although with larger scatter per star.

\section{STELLAR POPULATIONS OF $\omega$ CENTAURI}

\subsection{Color-Magnitude Diagrams}

Figure 4 shows the $V$, $B-V$ CMDs for each of the nine program fields, separately. 
The field F5 corresponds to the one placed on the cluster center. Figure 5 shows the 
$V$, $B-V$ CMD of all observed stars in $\omega$ Cen, where more than 130,000 stars in our 
program fields are plotted.  To eliminate the field stars superimposed on the $\omega$ Cen 
sequences (see Sec. 3.2), we matched $BV$ photometric data with those from $Ca$ \& Str\"omgren 
$by$ photometry. Due to the lower signal-to-noise ratio of the $Ca$ \& Str\"omgren $by$ data 
compared to that for $BV$, the matched CMD includes only about 85,000 stars. These color-magnitude 
data for $\omega$ Cen stars and for field star candidates with $V$ $<$ 16.0 are tabulated 
in Table 2 and Table 3, respectively. The x-y positions in the tables refer to the pixel 
coordinates relative to the origin of the field F1. The full data table will be published 
in a forthcoming paper. For the following analyses, we selected stars with a frame-to-frame 
errors in the $B$ and $V$ magnitudes of less than 0.05 mag, with the CMD resulting from this 
more restrictive cut containing only 38,000 objects.

\subsection {Field Star Elimination}

As shown in the CMDs, the signature of the Galactic field star contamination is evident, 
primarily as a swathe of stars with 0.4 $<$ $B-V$ $<$ 0.8 
for $V$ $<$ 16 and with many fainter redder stars. This field star 
sequence, which probably corresponds to the old disk stars (Jonch-Sorensen \& Knude 1994; 
Ng et al. 1995; Kaluzny et al. 1996, 1997), obviously complicates the interpretation of 
the data for some interesting sequences such as the red horizontal-branch (RHB), 
the asymptotic giant-branch (AGB), and the RGB.

The field star contamination can be minimized by using the metallicity-sensitive $hk$ index, 
defined as $(Ca - b) - (b - y)$ from the $Ca$ \& Str\"omgren $by$ photometric system 
(Anthony-Twarog et al. 1991; Twarog \& Anthony-Twarog 1995; Anthony-Twarog \& Twarog 1998). 
The $Ca$ \& Str\"omgren $by$ photometric system is an extension of the standard Str\"omgren 
$uvby$ intermediate-band photometric system with the inclusion of a fifth filter, $Ca$, 
centered on the K and H lines of Ca II (90 $\AA$ FWHM). We isolated relatively metal-rich 
field stars from the more metal-poor cluster members using the $hk$ vs. $b-y$ diagram.
 Because even the most metal-rich 
([Fe/H] $\sim$ -0.6, Pancino et al. 2002) stars in $\omega$ Cen are relatively metal-deficient 
compared to the typical disk field stars (Jonch-Sorensen \& Knude 1994; 
Twarog \& Anthony-Twarog 1995; Wyse \& Gilmore 1995), it is possible to eliminate foreground 
field stars through the use of the $hk$ index without seriously biasing the direct estimate 
of the abundance of stellar populations in the cluster.

As shown in the $upper$ $panel$ of Figure 6, at a given $b-y$, more metal-rich stars are 
offset from the metal-poor ones by virtue of a larger $hk$ index, therefore an upper envelope 
line which differentiates field stars can be drawn. Any star which lies below this envelope is 
tagged as a field star. The upper envelope of these field star candidates appears to cover 
empirical isometallicity lines for main-sequence dwarfs with [Fe/H] = -0.5 and 0.0 (solid lines, 
Anthony-Twarog et al. 2000). We selected 293 field star candidates for $V$ $<$ 16.0. 
The selected field stars show a prominent sequence, as expected, running through the middle 
of the $\omega$ Cen CMD (closed circles in the $lower$ $panel$ of Fig. 6). The foreground field 
stars at $V$ $>$ 16.0 are overlapped with the SGB and the RGB stars in the $hk$ vs. $b-y$ diagram making 
their identification difficult from photometry alone. The radial distribution of the field star 
candidates is uniform and is significantly different from that of the $\omega$ Cen RGB sequence, 
which confirms that most of our selected field star candidates are in fact not members of 
$\omega$ Cen (see Fig. 14a). This suggests that the use of the $hk$ vs. $b-y$ diagram is ideal for 
eliminating more metal-rich field stars from the member stars of $\omega$ Cen as well as most 
GGCs with reasonable resolution, and hence for constructing ``cleaned" CMDs. The $hk$ index is 
about 3 times more sensitive to metallicity than the $m_{1}$ index (see Fig. 9 of 
Twarog \& Anthony-Twarog 1995), therefore it is a better discriminator of field stars from cluster 
member stars. In Figure 7, we present the result of the decontamination of the CMD for $\omega$ Cen, 
showing data with frame-to-frame errors in the $B$ and $V$ magnitudes of less than 
0.05 mag after subtraction of the subset of field stars obtained from Fig. 6. 
The resulting ``cleaned" CMD has allowed us to obtain a more accurate distributions of bright 
stars on the RHB, AGB, and RGB
\footnote{
Note that the cleaning of the field stars in our study is valid only for $V$ $<$ 16.0 
and [Fe/H] $\geq$ -0.6 dex.
}.

\subsection {Multiple Stellar Populations from Red-Giant Branches}

One of the most notable features in our CMDs is the most metal-rich (MMR) sequence well separated 
from the bulk of bluer, more metal-poor stars, which was first discovered by Lee et al. (1999). 
The distinct feature of the MMR RGB sequence was later confirmed by Pancino et al. (2000) from 
their independent wide-field survey and could also be identified from other following surveys 
based on different photometric systems (e.g., Str\"omgren system of Hilker \& Richtler 2000; 
Washington system of Majewski et al. 2000; Frinchaboy et al. 2002).

In order to investigate the nature of the RGB, in Figure 8, we plot histograms of the RGB color 
distribution for two magnitude ranges (12.5 $<$ $V$ $<$ 13.0 and 13.0 $<$ $V$ $<$ 13.5).
Dereddened colors were derived, adopting $E(B-V)$ = 0.12 (Rey et al. 2000). The RGB stars are 
selected in a relatively  narrow magnitude range, so that the field star and the RGB bump 
contamination is minimized, and which also avoids mixing of the RGB stars with different 
metallicities in the histogram. The presence of three pronounced RGBs 
[i.e., most metal-poor (MMP), metal-poor (MP), and metal-rich (MR) RGBs] can be identified, 
while the MMR component is not as well distinguished as it is in the CMDs due to the small sample size 
of such stars in these magnitude ranges.

Because the intrinsic color of the RGB will vary with its metallicity, we obtained mean [Fe/H] 
abundances for the four main components using the relation between [Fe/H] and $(B-V)_{o}$ 
at fixed magnitude, based on the data of GGCs compiled by Ferraro et al. (1999a, hereafter F99).
The peaks in the histograms correspond to colors of fiducial lines of four components at
$V$ = 12.75 and $V$ = 13.25, which were adjusted by eye. From RGB fiducial lines of 
55 GGCs in the absolute plane of [$M_{V}$, $(B-V)_{o}$], which was kindly provided by 
F. R. Ferraro, we plotted the $(B-V)_{o}$ of the RGB measured at $M_{V}$ = -1.35 and -0.85 
as a function of [Fe/H] in the Zinn \& West (1984) scale. These two values of $M_{V}$ 
correspond to the mean magnitude of each magnitude range of Fig. 8, adopting $(m - M)$$_{V}$ = 14.1 
(Rey et al. 2000). Finally, we derived relations of [Fe/H] vs. $(B-V)_{o}$ for GGCs by 
third order polynomial fit (see also F99). We estimated their mean [Fe/H] abundances as 
$[Fe/H]_{ZW}$ $\sim$ -1.75, -1.45, -1.00, and -0.55, respectively, which are averaged from 
values obtained at two magnitude ranges using these relations. In Fig. 8, we label
the four main components and indicate their mean [Fe/H] abundances
\footnote{
We compared photometric errors with the dispersions of Gaussian fitting for four RGB components 
in the histograms, and found that the dispersions, which must arise from an intrinsic
metallicity spread, are much larger (factor of 2 - 5) than the photometric errors. 
}. 
The [Fe/H] values of the first three components agree well with those by Pancino et al. (2000)
at [Fe/H] = -1.7, -1.3, and -0.8, and by Frinchaboy et al. (2002) at [Fe/H] = -1.6, -1.2,
and -0.9. Our metallicity estimation for the MMR RGB component ([Fe/H] $\sim$ -0.55)
is also consistent with the value ([Fe/H] = -0.60) found by Pancino et al. (2002) 
from their high resolution spectroscopy observations for these stars 
\footnote{
Because the color of the RGB also slightly depends on the age of the stars, we have to consider an 
effect of different age of the MMR RGB stars compared to the dominant metal-poor stars. Assuming the
age of the MMP population is 12 Gyr and the MMR population is 4 Gyr younger than the MMP one 
(see Sec. 5), we calculate the offset in the RGB color associated with an offset in age as 
$B-V$ $\sim$ 0.05 mag at $M_{V}$ = -1 using the $Y^{2}$ isochrones with z = 0.005 (Yi et al. 2001). 
This produces only about 0.1 dex offset in [Fe/H] of the MMR component (i.e., [Fe/H] $\sim$ -0.45).
}.

In the light of the detection of four separate populations with different metallicities,
we divided our photometry of the RGB stars into four different sub-populations (i.e., MMP, MP, 
MR, and MMR RGBs). This division of the RGB stars will be adopted for the following 
analyses of the RGB stars. We relied on the shape of isochrones to divide the RGB stars into four 
sub-populations directly in the CMD, assuming the shape of the cluster RGB with different 
metallicity can be adequately represented by isochrones. We selected $Y^{2}$ isochrones 
(Yi et al. 2001), which go through the dips in the histogrms and fully cover the RGB color 
distribution of $\omega$ Cen. We fitted isochrones to quadratric or cubic relations and then, 
using these relations, divided all RGBs into four sub-populations in the plane of $M_{V}$ vs. $(B-V)_{o}$. 
The metallicity for each sub-population was empirically estimated using the relation 
of [Fe/H] vs. $(B-V)_{o}$ of GGCs, as explained above. We obtained relations between
[Fe/H] vs. $(B-V)_{o}$ at $M_{V}$ = -1, 0, and +1. The metallicity for each RGB sub-population
was obtained at each magnitude,  and then averaged. The mean [Fe/H] of the four sub-populations 
are $[Fe/H]_{ZW}$ $\sim$ -1.8, -1.5, -1.1, and -0.5, respectively, and their uncertainties are 
0.1 - 0.3 dex. These metallicities are, within the error, consistent with the ones obtained from 
the histogram of the color distribution (see Fig. 8). 

Figure 9 shows a zoomed CMD of the RGB region which is divided into four RGB sub-populations. 
In order to estimate the reliability of our division of the RGB stars into four sub-populations, 
and of their estimated metallicities, we compared them with [Fe/H] values ([Fe/H]$_{SK96}$) for 
a set of giants (bright giants and subgiants), which were obtained from spectroscopic Ca 
abundances by Suntzeff \& Kraft (1996, hereafter SK96). In Fig. 9, we have marked the common 
stars observed by SK96 and us using different symbols for different metallicity groups: closed circles 
are stars with [Fe/H]$_{SK96}$ $<$ -1.7, crosses have -1.6 $<$ [Fe/H]$_{SK96}$ $<$ -1.3, 
closed squares have -1.3 $<$ [Fe/H]$_{SK96}$ $<$ -0.7, and triangles have [Fe/H]$_{SK96}$ $>$ -0.7. 
The giants belonging  to the first three metal-poor groups of the SK96 sample are located consistently in 
our RGB sequences. However, the majority of the most metal-rich giants of SK96 deviate from 
the sequence of our MMR RGB. Most of these giants have [Fe/H]$_{SK96}$ $\sim$ -0.3 - -0.6, 
and therefore these should be located near our MMR RGB sequence, if our method dividing RGB 
sub-populations is reliable. It is worth noting that the Ca triplet calibrations adopted by 
SK96 are usually rather uncertain in the high metallicity regime (see Norris et al. 1996; 
Pacino et al. 2002). Furthermore, from the first direct abundance determination based on 
high-resolution spectroscopy for the most metal-rich RGBs (i.e., ``RGB-a" stars), 
Pancino et al. (2002) obtained their mean metallicity as $<$[Fe/H]$>$ $\sim$ -0.60, 
which is about +0.3 dex more metal poor than one obtained from SK96 (see also Origlia et al. 2003). 
While SK96 obtained [Fe/H] = -0.23 and -0.33 for ROA 300 and ROA 371, respectively, 
Pancino et al. (2002) present their metallicities as [Fe/H] = -0.77 and -0.95, respectively, 
which are more consistent with their locations in the CMD. In any case, we believe this 
ambiguity of metallicities for a few of the brighter MMR RGB stars will not invalidate 
our below analyses.

\subsection {Red-Giant Branch Bumps}

Most CMDs of GGCs show the intriguing feature of an RGB bump, which is an evolutionary pause 
on the first-ascent RGB (Thomas 1967; Iben 1968; Fusi Pecci et al. 1990; Cassisi \& Salaris 1997). 
The number of observational detections of the RGB bumps for GGCs has grown recently and the dependence 
of the RGB bump luminosities on their metallicities is now well established (Fusi Pecci et al. 1990;
Cassisi \& Salaris 1997; Alves \& Sarajedini 1999; F99; Zoccali et al. 1999). The RGB bump
of most GGCs generally is clumped due to the homogeneous metallicity among their stars.
However, considering the wide spread of metallicity of the $\omega$ Cen RGB stars, one would 
expect a continuous and slanting RGB bump in the CMD of $\omega$ Cen, because the luminosity 
of the RGB bump becomes fainter as its metallicity increases. The large sample of stars, 
coupled with the high photometric accuracy, allowed us to identify this feature on the RGB of 
$\omega$ Cen. As one can see in our CMDs (e.g., Fig. 8), from [($B-V$)$_{o}$, $V$] $\sim$ 
(0.8, 14.3) to (0.9, 14.9), an inclined sequence of the RGB bump is evident. The RGB bump of 
the MMR population with [Fe/H] $\sim$ -0.5 is less clearly defined, as its sample is not populous,
but probably is separated from the main feature of the RGB bump in the region of 
[($B-V$)$_{o}$, $V$] $\sim$ (0.95, 15.5).

In order to see the shape of the RGB bump in detail, we arbitrarily sliced the RGB into seven groups with 
different metallicities, and then examined their RGB bump features in the differential LFs.
In the $upper$ $panel$ of Figure 10, we present seven RGB groups. The division of different 
RGB groups and the determination of their metallicities (see Table 4) are the same as ones 
presented in Sec. 3.3. The LFs for seven RGB groups are presented in the $lower$ $panel$ of 
Fig. 10. These LFs were obtained by counting the number of stars in bins of 0.05 mag width 
in $V$. From the detection of slope change in the cumulative LF for each RGB group, 
we determined the magnitude of these RGB bumps. All bumps for the seven RGB groups are clearly 
defined and marked with arrows.

From our determination of luminosity and metallicity for seven RGB bumps of $\omega$ Cen,
we constructed a plot of the RGB bump luminosity against metallicity and compared this with a
general relation among GGCs, in order to see whether sub-populations in $\omega$ Cen follow
this general trend. F99 gave a tight relation for the RGB bump luminosity ($\Delta V_{ZAHB}^{bump}$),
in terms of the magnitude difference between the RGB bump ($V_{bump}$) and the zero age horizontal-branch 
($V_{ZAHB}$) in the RR Lyrae instability strip, as a function of metallicity using the data for
47 GGCs. In measuring the $\Delta V_{ZAHB}^{bump}$ of $\omega$ Cen, we follow the procedures 
consistent with those of F99. We adopted a single mean magnitude of the HB stars for all RGB bumps, excluding 
the most metal-rich one, as $<V_{HB}>$ = 14.64$\pm$0.10, which corresponds to the mean value for
RR Lyrae stars at [Fe/H] $\sim$ -1.5 (see Fig. 6b of Rey et al. 2000). It is difficult to select the 
MMR RHB clump stars, because the expected location of the RHB clump overlaps with that of the more
metal-poor RGB stars in the CMD. Therefore, from our synthetic HB models (see Sec. 5.1), we estimated 
$<V_{HB}>$ = 14.70$\pm$0.04 for the mean magnitude of the RHB clump with [Fe/H] $\sim$ -0.5 and 
age $\sim$ 8 Gyr. It should be noted that metal-rich clusters with [Fe/H] $>$ -1.0 have no RR Lyrae 
stars and their RHB magnitude would be brighter than that of the RR Lyrae stars, if some existed 
(see the discussion in Lee, Demarque, \& Zinn 1987, 1990; Castellani, Chieffi \& Pulone 1991). 
In order to provide a consistent determination of $<V_{HB}>$ with those of the metal-poor HB, 
the $<V_{HB}>$ of the RHB clump at the level of the RR Lyrae instability strip was evaluated by 
adding 0.10$\pm$0.05 mag (i.e., $<V_{HB}>$ = 14.80), according to the suggestion by 
Fullton et al. (1995) and Cassisi \& Salaris (1997). The $<V_{HB}>$ was converted to $V_{ZAHB}$ 
magnitude as suggested by F99 (see their Equation 2), and then finally, the $\Delta V_{ZAHB}^{bump}$ 
were derived. The $\Delta V_{ZAHB}^{bump}$ errors have been calculated by quadratically combining 
the errors on $V_{ZAHB}$ and $V_{bump}$, where the uncertainties of $<V_{HB}>$ and $V_{bump}$ are 
standard deviations of the mean and binning size of LF, respectively. Table 4 lists the relevant 
observational quantities for the RGB bumps of $\omega$ Cen.

In Figure 11a, we present the parameter $\Delta V_{ZAHB}^{bump}$ as a function of metallicity 
for the seven groups of the $\omega$ Cen RGB (closed circles). The data for 42 GGCs, is also plotted 
as open circles. The best fit to the data for GGCs obtained by F99 (see their Table 6) is also 
overplotted in the figure (dotted line). As shown in the figure, the behavior of $\omega$ Cen 
regarding its relation between luminosity and metallicity of the RGB bumps shows excellent agreement 
with that for the GGCs. In order to avoid an uncertainty of ZAHB level of the MMR population predicted 
by our synthetic HB models, it is useful to see directly the absolute magnitude of the RGB bump [$M_{V}(bump)$] 
as a function of metallicity. In Figure 11b, we plotted $M_{V}(bump)$ against [Fe/H]$_{ZW}$ for 
42 GGCs from the catalogue of F99 and superimposed a best fit to the data as a dotted line,
which shows a tight correlation. The RGB bumps with different metallicities of $\omega$ Cen
also follow the general relation found for GGCs in this diagram.

\subsection {Horizontal-Branch Stars}
From the distinct multiple populations of the $\omega$ Cen RGB stars and their wide spread of 
metallicity, it is naturally expected that the HB distribution of $\omega$ Cen should be 
represented by a mixture of different populations following the internal age-metallicity relation 
(Lee et al. 1999; Hughes \& Wallerstein 2000; Hilker \& Richtler 2000; see Sec. 5 for details).
Figure 12 shows the HB region of the CMD after subtraction of the field stars. In this figure, 
the RR Lyrae variables are plotted using mean $V$ and $B-V$ values from our photometric data. 
The HB of $\omega$ Cen is predominantly blueward of the RR Lyrae instability strip, which is 
in agreement with the low metallicity of the majority population of this cluster. It is also 
clear that the BHB of $\omega$ Cen shows a long tail of very blue stars, with gaps occurring 
at $V$ $\sim$ 16.6, 17.0, and 17.7. The overall HB distribution is consistent with the discrete 
nature of the RGB populations: there are BHB and RR Lyrae variables mainly associated with the 
metal-poor and old RGBs (i.e., MMP and MP RGBs), some RR Lyrae and RHB stars associated with 
the metal-rich and slightly younger population (i.e., MR RGB), and finally the RHB clump 
superimposed on the RGB, which must be associated with the most metal-rich and younger 
population (i.e., MMR RGB; see Sec. 5 for details).

The ``cleaned" CMD has allowed us to obtain a more accurate distribution of the RHB stars which 
must be located from the red edge of the RR Lyrae instability strip to the base of the AGB stars. 
We defined RHB stars as having 14.20 $<$ $V$ $<$ 14.85 and  0.55 $<$ $B-V$ $<$ 0.71, marked by 
a box (dashed line) in the CMD. The CMD indicates that the RHB stars are loosely grouped and 
separated by a luminosity gap from stars at the base of the AGB. The vertical extent of the 
RHB stars in the CMD is about 0.7 mag. This indicates that the majority of these stars are 
metal-rich ([Fe/H] $\sim$ -1.0) and have higher stellar masses, following blue loops prior to 
rapidly evolving redwards (Lee \& Demarque 1990; Yi, Demarque, \& Kim 1997). Some bright RHB 
stars are probably not ZAHB stars, but may be MP ([Fe/H] $\sim$ -1.5) stars that have evolved 
from the BHB, at the final phase of core helium burning, to the AGB (Lee et al. 1990). Several 
bright stars, about 0.7 mag above the mean magnitude of the RHB stars, with 0.58 $< B-V <$ 0.73 
and 13.5 $< V <$ 14.0 are believed to be highly evolved low mass stars from the MMP 
([Fe/H] $\sim$ -1.8) BHB. There is a group (13.95 $<$ $V$ $<$ 14.60, 0.71 $<$ $B-V$ $<$ 0.79) 
of stars slightly brighter than the RHB stars which are well separated from the base of the AGB 
but mixed with stars in the red edge of the RHB. They are probably the BSS progeny candidates evolved from the main population 
of the BSS (see Sec. 3.6).

\subsection {Blue Straggler Stars}

Due to small sample size and large photometric errors, only a few BSSs have been reported from
the early photometry for $\omega$ Cen (e.g., Da Costa, Norris, \& Villumsen 1986). Recently, 
Kaluzny et al. (1996, 1997) reported about 200 BSS candidates as a byproduct of their OGLE project. 
Our photometry has also revealed a significant population of the BSS in $\omega$ Cen. In our CMDs, 
most of the BSSs form a well defined sequence extending from the MSTO to about 2 mag brighter 
than the MSTO.

Fig. 12 shows the BSS region of the CMD where the total of 222 BSSs are plotted as closed circles. 
The definition of the ``BSS region'' is somewhat arbitrary at the edge between the BSS region and 
the upper MS. It should be noted that, unlike the distribution of the BSSs in many other GCs, 
the blue edge of the BSS sequence of $\omega$ Cen is clearly defined. We shall refer to it 
as the blue straggler main sequence (BSMS) following the suggestion of Mandushev et al. (1997) 
from a narrow and tight BSS distribution seen in M55. The BSS sequence redder than the BSMS shows 
a broad distribution in color. The spread in color is probably intrinsic and could be attributed 
to the effect of stellar evolution of the BSSs with different masses off the BSMS. The broadening may 
partially result from the wide spread of metallicity of $\omega$ Cen, given that the mean location 
of the cluster BSS sequence is correlated with its metallicity, since the BSS sequences are 
progressively located to the redside with increasing cluster metallicity (Sarajedini \& Da Costa 1991; 
Fusi Pecci et al. 1992). However, considering the fact that the metal-poor population comprises 
more than 80 \% of the $\omega$ Cen population, this metallicity effect may be small.

In Fig. 12, we overplot $Y^{2}$ isochrones (Yi et al. 2001) for the ZAMS (solid line) and 
a few different ages (2, 3, 4, 5, and 6 Gyr, long dashed lines) for the BSS region of the CMD.
These isochrones match the MMP population of $\omega$ Cen with Z = 0.0006 (see Sec. 5), and 
correspond to a MSTO mass range of about 1.0 - 1.5 M$_{\odot}$, consistent with the general 
expectation of higher masses for the BSSs than for the MS stars. We also plot the 12 Gyr isochrone 
(gray line) as a reference. The brightest BSS lies below the 2 Gyr isochrone, which suggests that 
the masses of the BSSs in our sample do not exceed 1.4 M$_{\odot}$. In our final list of the BSSs, 
we excluded a few bright stars which are located above the BSS sequence at $V$ $\sim$ 15.2 - 15.8, 
due to their large scatter from the main BSS sequence and their proximity to the HB. 

There is an intriguing group of stars (crosses in Fig. 12) which are separated from the 
bulk of the main branch of the BSSs and lie above the cluster SGB in the region of 
16.0 $<$ $V$ $<$ 16.8 and 0.40 $<$ $B-V$ $<$ 0.55. At first glance, these stars appear to 
lie in a series of distinct sequences evolving from the BSSs towards redder colors. 
They are unlikely to be field stars, because, at $V$ $\sim$ 16.5, they have bluer 
$B-V$ colors than expected for the foreground field sequence. Such stars, so called 
``yellow straggler stars" (YSSs), have been found in many clusters, and interpreted as 
being an optical blend of BSSs with SGB stars (Ferraro, Fusi Pecci, \& Buonanno 1992; 
Ferraro et al. 1992). The images of these stars were inspected on the CCD frames 
in order to check the possibility that our photometry of these stars has been affected 
by their neighboring stars. Most (about 70 $\%$) of these stars are found to be isolated 
and clearly defined. Hesser et al. (1984) explained the YSSs as being evolved BSSs which are 
located between the BSSs and the cluster giant branch (e.g., post-mass-transfer binaries, 
Landsman et al. 1997). The lack of YSSs in many GCs is ascribed to the effects of rapid 
stellar evolution after the hydrogen exhaustion in the core of the BSS 
(Sarajedini \& Da Costa 1991; Sandquist, Bolte, \& Hernquist 1997). However, considering 
the large number of  BSSs in $\omega$ Cen, we cannot rule out a possibility that some 
of these stars are chemically unmixed structures evolved from the main BSS sequence
(e.g., Sandquist et al. 1997). On the other hand, these YSSs appear to be separated from 
the main BSSs by a gap and are located about 0.7 mag brighter than the BSSs.  
This suggests that, alternatively, some of these stars might be the binary BSSs with 
roughly equal masses (Mandushev et al. 1997). This latter hypothesis is consistent with 
the observed centrally concentrated radial distribution of the YSSs (see Sec. 4.3) 
and the hint of higher probability of finding binary BSSs in the inner region of $\omega$ Cen 
(see Table 8 of Kaluzny et al. 1997).

It is suggested that some stars located both redder and brighter than the normal RHB stars
could be the BSS progeny candidates, which may be evolved from the BSSs and are in the core 
helium burning  phase (Renzini \& Fusi Pecci 1988; Fusi Pecci et al. 1992). In several GGCs, 
a few BSS progeny candidates were identified (Fusi Pecci et al. 1992; Ferraro et al. 1997a, 
1997b, 1997c, 1999b). Given our finding of a large population of the BSSs and a few YSSs, 
we can also expect the existence of the BSS progeny candidates in $\omega$ Cen. 
From the comparison with a CMD of M3 (Ferraro et al. 1997a), we estimate the possible region 
of the $\omega$ Cen BSS progeny stars in our CMD. In Fig. 12, we define the $\omega$ Cen BSS 
progeny region as a box (solid line; 13.95 $<V<$ 14.60, 0.71 $<B-V<$ 0.79) following that of 
M3 represented by Ferraro et al. (1997a, see their Fig. 18). There are 24 stars in this region, 
which are slightly brighter than the normal RHB stars and well separated from the base of the AGB.

According to the theoretical models for the formation of BSSs, the BSS LF may be a good tool
to discriminate between different origin scenarios: stellar collisions and primordial binary
mergers (e.g., Bailyn \& Pinsonneault 1995, hearafter BP95). The BSSs formed by two different 
processes would lead to different lifetimes and evolutionary tracks in the CMD and, therefore,
lead to different LFs for the BSSs. The BSSs created by primordial binary mergers would be seen 
less often at the bright end of the LF compared to the BSSs created by stellar collisions. 
As shown in Figure 13, we obtained observed BSS LFs of $\omega$ Cen for the inner ($<$ 3r$_{c}$, 
$upper$ $panel$) and outer regions ($>$ 3r$_{c}$, $lower$ $panel$) of the cluster, respectively, 
and compared those with the models of BP95. The error bars represent Poisson noise. The models 
of BP95 are represented by a dashed line (collisional model) and a dotted line (primordial binary 
merger model), respectively. Estimating the MSTO magnitude of $\omega$ Cen as $V_{MSTO}$ = 18.2 
from our photometry and assuming a differential bolometric correction of -0.1 mag between the BSS 
and the MSTO stars, we derive the BSS bolometric luminosity from the relation $log(L/L_{MSTO}) = 0.4(V_{MSTO} - V - 0.1)$, following BP95. The observed LFs of the inner and outer regions appear 
to be different from each other, at the 99.9 \% confidence level from the Kolmogorov-Smirnov (K-S) 
test. While the BSSs in the inner region show a flat LF for all magnitudes, the BSS LF of the outer 
region has a shape of increasing numbers of stars with magnitude. Furthermore, the shape of the 
BSS LF of the inner region appears to be similar to that of the collisional model (dashed line), 
while the LF of the outer region resembles the merger model (dotted line). The comparison 
in Fig. 13 presents observational support for the hypothesis that the origin of BSSs might be 
at least partially affected by overall environmental conditions even in different regions 
of a single GC (Ferraro et al. 1997b; Rey et al. 1998), in addition to the case of GCs with 
different overall structural properties (Fusi Pecci et al. 1992; Sarajedini 1993; Piotto et al. 1999).

\section{RADIAL DISTRIBUTIONS}

\subsection {Red Giant Branch stars}

Norris et al. (1997) noted that the metal-rich RGB stars of $\omega$ Cen are more 
centrally concentrated than the metal-poor ones. From their wide-field observations, 
Hilker \& Richtler (2000) also confirmed that the metal-rich RGBs with [Fe/H] $\sim$ -0.9 
are more centrally concentrated than the metal-poor counterpart with [Fe/H] $\sim$ -1.8. 
However, they found that the most metal-rich RGB stars with [Fe/H] $>$ -0.5 appear less 
concentrated than the other metal-poor populations. On the other hand, Pancino et al. (2000, 2003) 
suggested that their most metal-rich RGB stars (i.e., ``RGB-a" stars) have a similar 
distribution with their intermediate metallicy RGB ones, and both seem to be more centrally 
concentrated than the most metal-poor population.  

The cumulative radial distributions of our four RGB sub-populations (see Sec. 3.3) 
with different metallicities are plotted as a function of the projected radius in Figure 14a. 
We included only bright RGB stars with $V$ $<$ 15.0 for these distributions. As a reference, 
we present a distribution of the probable foreground field stars (thick gray line) extracted from 
our $Ca$ \& Str\"omgren $by$ photometry (see Sec. 3.2). The radial distribution of the selected 
field star candidates is uniform, and significantly different from those for the $\omega$ Cen 
RGB stars. Two-sided K-S tests show that the radial disribution of the MP RGB is different from 
that of the MPP and the MR RGB at the 77.4 \% and 57.2 \% confidence level, respectively. 
The radial distribution of the MR RGB is found to be different from that of the MPP RGB
at the 96.9 \% confidence level. However, we could not confirm a central 
concentration of more metal-rich RGB stars compared to metal-poor counterparts as suggested 
by previous studies (Norris et al. 1997; Hilker \& Richtler 2000; Pancino et al. 2000).

A notable feature of Fig. 14a is the radial distribution of the MMR RGB with 
[Fe/H] $\sim$ -0.5 which is different from those of other more metal-poor ones. 
The MMR RGB appears to marginally show a bimodal distribution. The cumulative radial 
distribution suggests that the MMR RGB has a slight dip in the distribution between 
$r$ $\sim$ 8$\arcmin$ and $r$ $\sim$ 17$\arcmin$. A two-sided K-S test applied to two 
distributions between the MMP and the MMR RGBs confirms that they are different at the 
99.6\% significance level. Since all RGB samples have approximately the same photometric 
accuracy and completeness, our analysis tentatively suggests that the MMR RGB stars have 
a bimodal radial distribution, with a high frequency in the inner and outer regions but 
a dip in the intermediate region. In Figure 15, we plot a x-y position map for the MMP 
(Fig. 15a) and the MMR (Fig. 15b) RGB stars with $V$ $<$ 15.0.  A gap of the distribution 
in the  MMR RGB stars toward the east direction, which can primarily explain this bimodal 
distribution, is visible. On the other hand, the MMP RGB stars show a different distribution 
without a hint of bimodality. For more clarity, in Fig. 14b and Fig. 14c, we also present 
cumulative radial distributions of the MMP and the MMR RGB stars in the regions of east 
and west direction, respectively. The K-S tests show that the bimodal distribution of 
the MMR RGB in the eastern region appears to be more significant (99.9\%) than that for 
the western region (90.1 \%). A group of the MMR RGB stars in the outer part of the 
eastern region seems to be isolated from the main distribution, although statistical 
significance is low due to small number counts for these stars. However, recently, 
Pancino et al. (2000, 2003) also presented a complex and perturbed distribution for 
their most metal-rich RGB stars [see also Fig. 3 of Pancino et al. (2000) and Fig. 1 
of Pancino et al. (2003), respectively]. Both our and Pancino's results suggest that 
there might be a peculiar substructure among the MMR RGB stars of $\omega$ Cen.

\subsection{Horizontal Branch Stars}
We present the radial distributions for the BHB and the RHB stars in Figure 16. In order to check 
if they are consistent with being drawn from the same parent distribution of the RGB, we also 
present radial distributions of two RGB populations (gray lines) with [Fe/H] $\sim$ -2.0 - -1.3 
(i.e., MMP and MP RGBs) and [Fe/H] $\sim$ -1.0 (i.e., MR RGB) which must be associated with 
the BHB and the RHB, respectively. We have made the comparison for stars with a roughly equivalent 
magnitude range of 14.0 $< V <$ 15.0, so as to minimize a difference of completeness between 
the HB and the RGB stars. In Fig. 16a, the BHB shows a consistent distribution with that of the MMP 
and the MP RGB populations. In contrast, the radial distribution of the RHB is somewhat different 
from that of the correspondng MR RGB (Fig. 16b), with the RHB (solid line) appearing to be more 
centrally concentrated than the MR RGB
\footnote{
It is difficult to find a reason for the inconsistency between the  RHB  and the MR RGB radial
distributions. From an evident feature of the field star sequence extending through the RHB in the 
CMD (see lower panel of Fig. 6), it might be suggested that some field stars from our selection 
could really be bona-fide RHB stars, if our field star selection was overly restrictive. 
In this case, if we convert some field stars, whose radial distribution is uniform for all radii 
(see Fig. 14a), into RHB stars, the radial distribution of the RHB can be made similar to that of 
the MR RGB. However, we consider this explanation rather unlikely, considering the clear separation between the RHB and the field star candiates in the $hk$ vs. $b-y$ distribution (see upper panel of 
Fig. 6).
}.
A one-sided K-S test shows that the distribution of the RHB stars is different from that of 
the MR RGB stars at the 77.6\% confidence level. In Fig. 16c, we also compared the radial 
distributions between the BHB and the RHB. The RHB shows more central concentration than the BHB 
with 89.2\% K-S probability. This result is consistent with the situation for most dwarf 
galaxies in the local group (Harbeck et al. 2001). 

We show a radial distribution of stars located in the hot and faint HB (16.0$<V<$18.4, 
-0.40$<B-V<$0.15) in Figure 17. For comparison, we also plot a distribution of the MS and 
SGB stars (16.0$<V<$18.4, 0.5$<B-V<$0.8) in the same magnitude range of the hot HB stars.
The MS and SGB stars have been chosen to lie within $\pm$0.05 mag in $B-V$ of the cluster 
fiducial sequence 
\footnote{
Although the MS and SGB stars have been selected around the fiducial, a small contamination
by field stars might still be present for $V$ $>$ 16.
}.
The hot HB stars are apparently more concentrated toward the cluster 
center than are the MS and SGB stars. From one-sided K-S test, we found that the central 
concentration of the hot HB stars is more significant with a 99.0\% confidence level. 
This result is consistent with that of Bailyn et al. (1992) that extreme BHB stars of 
$\omega$ Cen are centrally concentrated with respect to the SGB and the BHB stars.

\subsection{Blue Straggler Stars}
We compare the radial distribution of the BSSs with that of the SGB/MS stars selected to have 
similar photometric uncertainty and completeness. The SGB/MS stars have been chosen to lie 
within $\pm$0.05 mag in $B-V$ of the cluster fiducial sequence between 16.0 $<$ $V$ $<$ 18.4.
Figure 18a shows the cumulative radial distribution of the BSSs (solid line) compared with 
the SGB/MS stars (dotted line). This figure appears to indicate that the radial distribution of 
the $\omega$ Cen BSSs is similar to that of the SGB/MS stars with similar magnitude and might seem to 
contradict the general results that the BSSs are more massive and more centrally concentrated than 
the other cluster stars, in the context of the BSS formation scenario as being due to the merger 
of close binary stars or to the direct collision of two or more stars (Stryker 1993; Bailyn 1995). 
In Fig. 18b, in order to secure more stars in the central region, we present cumulative radial 
distributions for the BSSs and the SGB/MS stars with frame-to-frame error in the $B$ and $V$ 
magnitudes of less than 0.1 mag, which is larger than that used for our final good photometric 
list and Fig. 18a. A K-S test applied to these two distributions indicates the BSSs to be more 
centrally concentrated than the SGB/MS stars at the 97.0\% level. Although our results are 
based on data with some large errors, we suggest it can not be ruled out that the radial 
distribution of the BSSs is different to that of the SGB/MS stars, the former being more concentrated toward the cluster center. The spatial distribution of the YSSs is compared with that of the bright 
(16.0 $<$ $V$ $<$ 17.2) BSSs in Fig. 18c. The YSSs appear to be more centrally concentrated than 
the bright BSSs. A one-sided K-S test shows that the distribution of the YSSs is different from 
that of the bright BSSs at the 95.3\% significance level. This suggests that the YSSs are slightly 
higher mass single stars that evolved from the main BSS population, or alternatively, may be the high mass binary BSSs.

\section{AGE-METALLICITY RELATION OF $\omega$ CEN}

For sufficiently old ( $>$ 8 Gyr) stellar populations, evolutionary models (e.g., Lee et al. 1994)
predict that the surface temperature distribution of the HB stars is very sensitive to age, which makes
it an attractive age indicator from the interpretation of the HB morphology in the CMDs. For a fixed
metal abundance, age differences of several billion years produce dramatic changes in the morphology
of the HB in the sense that older clusters have bluer HB's. Our most recent update of the HB models
makes the HB morphology more sensitive to age (see Fig. 9 of Rey et al. 2001) compared to our
earlier models (Lee et al. 1994). For the case of clusters and dwarf galaxies with multiple stellar
populations and internal age-metallicity relation among them, once we have information on metallicities
of different populations it is possible to interpretate a HB morphology in terms of relative age
differences between different populations. In our following analysis using our updated synthetic 
HB models for $\omega$ Cen, we have estimated the age difference between four sub-populations
from the HB morphology once a good match is obtained between the observed and model HB's. In particular,
we have made an estimation of the total age spread within $\omega$ Cen from the location of 
the RHB clump (i.e., the MMR RHB) associated with the MMR RGB with respect to the BHB associated 
with the MMP RGB.

\subsection{Horizontal-Branch Models and Synthetic Color-Magnitude Diagram}

We constructed synthetic CMDs of $\omega$ Cen using our most updated version of the HB population
models based on the Y$^2$ isochrones (Yi et al. 2001) and corresponding HB evolutionary tracks
(Yi, Demarque, \& Kim 1997). First of all, we assumed the absolute age of the oldest GGCs and 
the MMP population of $\omega$ Cen is about 12 Gyr, as suggested by the Hipparcos distance 
calibration and other improvements in stellar models (Gratton et al. 1997; Reid 1998; 
Chaboyer et al. 1998; Grundahl, VandenBerg, \& Andersen 1998; Yi et al. 2001). We assumed primodial helium abundance $Y_{p}$ = 0.23 and a helium enrichment factor $\Delta$Y/$\Delta$Z = 2, respectively 
(see Yi et al. 2001). We adopted [$\alpha$/Fe] = 0.3 for the three metal-poor populations with 
[Fe/H] $<$ -0.8 (Norris \& Da Costa 1995; Smith et al. 2000), and we assumed that it declines to 
0.1 for the most metal-rich population of $\omega$ Cen (Pancino et al. 2002; Origlia et al. 2003).  
The treatment suggested by Salaris, Chieffi, \& Straniero (1993) was used to simulate the effect 
of $\alpha$-element enhancement. The empirical mass-loss law of Reimers (1975) was used in order to 
consider the effect of age and metallicity on the mass-loss from the RGB to the HB. The reader is
referred to Lee et al. (1990, 1994) and Park \& Lee (1997) for the details of model constructions 
(see also Rey et al. 2001).

After a series of model calculations with various combinations of metallicity and age, 
we found our best estimates for the metallicities and ages of four distinct sub-populations 
as [$Z$, Age (Gyr)] = (0.0006, 12), (0.001, 12), (0.002, 10.8), and (0.006, 8). 
The metallicities correspond to the [Fe/H] obtained from the peak values in the color distribution 
of the RGB stars (see Fig. 8). In Figure 19, the observed CMD of $\omega$ Cen (grey dots) is
overplotted by our synthetic HB models, which include the scatter expected from the random 
errors in magnitude and in color as estimated by our photometry. We also overplotted
four $Y^{2}$ isochrones representing four distinct populations. One can see from Fig. 19 that 
the HB morphology produced by our synthetic HB models is consistent with the observed complex 
nature of the RGB (see also Fig. 1 of Ree et al. 2002). The RHB clump associated with the MMR RGB 
overlaps in the CMD with the RGB bumps of the less metal-rich populations. Note again that 
the BHB is from the combination of the metal-poor and old populations (i.e., MMP and MP RGBs), 
the RHB is from the metal-rich population (i.e., MR RGB), and finally the RHB clump is from 
the most metal-rich and younger population (i.e., MMR RGB).

\subsection {Age-Metallicity Relation}

RR Lyrae stars are intrinsically an abundance-biased sample due to the failure of the
metal-rich RHB stars to penetrate into the instability strip. Furthermore, the frequency
of RR Lyrae stars at a given metallicity depends on the HB morphology as well as the
metallicity distribution of the underlying stellar population (e.g., Lee 1992 for 
RR Lyrae stars in Baade's window). Our models suggest that the RR Lyrae variables are
mainly produced by the two metal-poor (MMP and MP; $Z$ = 0.0006, 0.001) populations with 
similar ages ($\sim$ 12 Gyr), and some RR Lyraes are from the younger ($\sim$ 10.8 Gyr) 
metal-rich (MP; $Z$ = 0.002) population, which is consistent with the metallicity measurements 
of these stars (Butler et al. 1978; Gratton et al. 1986; Rey et al. 2000). For the RR Lyrae 
stars in $\omega$ Cen, we have new metallicity measurements from our $Ca$ \& Str\"omgren $by$
photometry (Rey et al. 2000), and therefore more detailed comparison with the models is
possible. We confirm that our models reproduce the luminosity vs. metallicity and 
the period-shift vs. metallicity relations of the RR Lyrae variables in $\omega$ Cen, 
including the sudden upturn of the RR Lyrae luminosity and corresponding increase in 
period-shift at [Fe/H] $\sim$ -1.5, which is a result of redward evolution from 
the BHB (see Rey et al. 2000; Yoon \& Lee 2002). This suggests we have a 
very good understanding of what is happening in the instability strip of $\omega$~Cen.

The estimated ages of the MP and the MR populations in our HB models are based on the observational
relations, such as luminosity vs. metallicity and period-shift vs. metallicity of the
RR Lyrae stars in $\omega$ Cen (Rey et al. 2000). The suggested age (12 Gyr) of the MP 
population is supported by the observed number of the RR Lyrae stars in the range of 
[Fe/H] $\sim$ -2.0 - -1.5. If the age of the MP population is younger than the MMP population 
by 1 Gyr, for example, more HB stars of the MP population could penetrate into the instabilty strip, 
and the number of the RR Lyrae stars would be increased, which is not consistent with the observational 
results. The HB color distribution of the MR component ($Z$ = 0.002) in our models is almost 
centered at the red edge ($B-V$ = 0.5) of the instability strip with same number of stars 
on either sides. The observational data suggest that some metal-rich RR Lyrae stars with 
[Fe/H] $\sim$ -1.0 do exist and they would be unevolved ZAHB stars (see Rey et al. 2000). 
With these constraints from the RR Lyrae distribution in $\omega$ Cen, we estimate the age 
of the MR population to be about 1.2 Gyr younger than the MMP and the MP populations
\footnote{
While the BHB can be reasonably fitted by HB models with metal-poor and old populations,
the model RHB stars ($\sim$ 80 stars) are too many to be explained by the number of observed stars 
($\sim$ 50 stars). Furthermore, the observed mean luminosity of the RHB stars is brighter 
than the expected ZAHB levels by about 0.2 mag. As shown in Fig. 12, we suggest some bright 
stars in the region of the RHB might be old metal-poor stars which have evolved from the BHB 
towards the AGB. Although there are some discrepancies between models and observations for 
the MR RHB stars, we strongly suggest that the MR RHB stars are  not more than $\sim$ 2 Gyr younger than 
the MMP population.
}.

Our models also reproduce $\sim$ 150 RHB clump stars associated with the MMR RGB stars, 
but only when the age of the MMR population is somewhat younger (a few Gyr) than the MMP 
population. This is illustrated in Figure 20, where we can see the variation of the RHB 
clump location in the CMD under different assumptions regarding the age difference between 
the MMP and the MMR populations. These figures illustrate the relative age estimation from 
the location of the RHB clump associated with the MMR RGB with respect to the BHB and the MMP 
RGB stars. Figure 20a shows the case where all the stars in $\omega$ Cen have the same age 
despite their different metallicities. However, from the shape of the LF of the MMP RGB 
(see the  LF of the RGB1 in Fig. 10), we can rule out the case of no age difference between the 
$\omega$ Cen populations, because the observed LF does not show a prominent peak at V $\sim$ 14.8 
which corresponds to the mean magnitude of the RHB clump predicted by our models. Furthermore, 
this predicts the blue part of the RHB clump should be clearly bluer than the MMP RGB sequence 
at the level of the HB. Age differences greater than 6 Gyr (Fig. 20d) can also be excluded 
from the observed LF of the RGB (RGB6 of Fig. 10), because the observed LF of the RGB  
does not show a significant peak at V $\sim$ 14.7 due to enhancement of the RGB by the RHB clump. 
On the other hand, the observed CMDs do not show a spur of the MMR MSTO stars rising above 
the MMP SGB stars, similar to the feature corresponding to the metal-rich and young 
($\Delta t$ $\sim$ 7 Gyr) population seen in the CMD of the Sagittarius dwarf galaxy 
(see Layden \& Sarajedini 2000). Therefore we conclude that the acceptable range of 
age differences are $\Delta t$ $\sim$ 2 - 4 Gyr as shown in Fig. 20b and c.

This conclusion is somewhat affected by the presence of the RGB bumps associated with 
the discrete RGBs (see Sec. 3.4), because the expected location of the RHB clump
overlaps with that of the more metal-poor RGB bump on the CMD. Therefore, it is important to
discriminate the RHB clump from the more metal-poor RGB bump stars for better estimation
of relative age between the MMP and the MMR populations. Figure 21 shows the CMD 
that highlights the RGB and the RGB bump region. For the purpose of the analysis, the RGB 
is divided into three groups with different metallicities. The RGB bump is shaped as an 
inclined sequence due to the metallicity spread of the $\omega$ Cen RGB stars. 
A smaller box indicates the predicted location of the RHB clump with [Fe/H] $\sim$ -0.5,
from our synthetic HB models, which is younger ($\Delta t$ $\sim$ 4 Gyr) than the MMP population. 
As shown in the $upper$ $panel$, if $\Delta t$ $\sim$ 4 Gyr, then the RHB clump is predicted 
to be superimposed on the RGB bump of the intermediate metallicity (-1.5 $<$ [Fe/H] $<$ -1.0) 
population, and therefore we could expect an enhancement of stars on the intermediate metallicity 
RGB by the RHB clump stars. In the $lower$ $panel$, we compared normalized LFs of the metal-poor 
([Fe/H] $<$ -1.5) RGB stars (dashed line) with that of the intermediate metallicity RGB ones 
(solid line). The LF of the metal-poor RGB is shifted horizontally in order to coincide its 
RGB bump with that of the intermediate metallicity RGB. An arrow indicates the predicted location 
of the RHB clump, from our models, which is younger ($\Delta t$ $\sim$ 4 Gyr) than the MMP population. 
It is worthwhile to note that the peak of the intermediate metallicity RGB bump appears to be 
higher than that of the metal-poor one, indicating that the RHB clump stars might enhance the 
intermediate metallicity RGB bump feature due to their similar locations. A one-sided K-S test 
shows that the peak of the LF of the intermediate metallicity RGB is higher than
that of the metal-poor one at the 99.5 \% significance level. In the case of 
$\Delta t$ $\sim$ 2 Gyr from our HB models, we did not find a significant enhancement 
of the LF of the RGB. From this and similar analyses under different assumptions regarding 
$\Delta t$, we conclude that a $\Delta t$ of $\sim$ 4 Gyrs is our best estimate for the age spread 
within $\omega$ Cen, in the sense that the MMR population is younger. 

Our conclusion is in qualitative agreement with the results based on the Str\"omgren photometry
of MS stars (Hughes \& Wallerstein 2000; Hilker \& Richtler 2000). Recently, from their 
preliminary results obtained from AAT 2dF observations for a sample redwards of the MS, 
Cannon et al. (2002) found that faint and red members of $\omega$ Cen, which must be the most
metal-rich and young population in the vicinity of the SGB, do indeed exist. In Table 5, 
we summarize our final suggestions on the age-metallicity relation and characteristics of the
four distinct populations in $\omega$ Cen

\section {DISCUSSION}
\subsection{The Origin of $\omega$~Centauri}

The recent discovery of multiple stellar populations and the internal age-metallicity
relation of $\omega$ Cen found in our work and others (Lee et al. 1999; Pancino et al. 2000;
Hilker \& Richtler 2000; Hughes \& Wallerstein 2000), resembling the characteristics of the
Sagittarius dwarf galaxy, suggest that $\omega$~Cen was massive enough for the processes of 
self-enrichment and several bursts of star formation to proceed over several Gyr. 
These properties of $\omega$ Cen are also consistent with the characteristics of Local Group 
dwarf spheroidal galaxies, which are known to have complex star formation histories with 
extended timescales, and clearly distinct from the single age and metallicity characteristics of 
a typical GGC (Mateo 1998; van den Bergh 1999; Grebel 2001).

The relatively extended enrichment period of $\sim$ 4 Gyrs then indicates that the initial 
evolution of $\omega$~Cen occurred away from the dense central regions of the young Milky Way 
because at smaller galactocentric distances, such as the current location of $\omega$~Cen, 
we would expect the gaseous materials to have been stripped from the proto $\omega$~Cen on a 
much shorter timescale as a result of disk shocking or similar processes. This view is not 
inconsistent with the rather unusual kinematics and retrograde orbit of $\omega$ Cen which would 
allow the proto $\omega$ Cen to have a long orbital decay time (Dinescu, Girard, \& van Altena 1999;
Dinescu 2002; Mizutani, Chiba, \& Sakamoto 2003).  This scenario also is supported by 
the interesting result that the s-process elements are enriched over the whole [Fe/H] range 
(Norris \& Da Costa 1995; Smith et al. 2000), which implies a sufficient time interval 
(order of 2 - 3 Gyr) to allow the formation of the low-mass (1 - 3 $M_{\odot}$) AGB stars from 
interstellar gas retained in the deep potential well.

From the apparently discrete nature of the RGB in the CMD as well as dynamical and structural 
pecularities between RGBs with different metallicities, it might be suggested that $\omega$~Cen 
could be a merger of two or several GCs (Norris et al. 1997; Smith et al. 2000; see also Catelan 1997). 
But mergers of clusters are very unlikely, if not impossible, in the Milky Way, considering the 
rather high velocities between the clusters in the halo. Furthermore, the broad metallicity 
distribution of the  $\omega$ Cen RGB cannot be explained by simple merging of monometallic 
GCs (Pancino et al. 2002). We cannot rule out this scenario in dwarf galaxies however, 
because the internal velocity dispersion is much lower than in the Milky Way (van den Bergh 1996; 
Thurl \& Johnston 2002; Fellhauer \& Kroupa 2003).

On the other hand, the origin and evolution of the MMR population with [Fe/H] $\sim$ -0.5
may play a key role in the picture of the formation and evolution of $\omega$ Cen.
Pancino et al. (2000) and Hilker \& Richtler (2000) found that the spatial distribution of 
the MMR population is significantly different from that of the MMP one. Furthermore, 
Norris et al. (1997) earlier reported that the metal-rich stars do not show rotation, 
whereas the metal-poor ones do rotate. These studies suggest that the MMR population may 
have a different formation mechanism with respect to the bulk of the cluster popuation. 
Within the hypothesis that $\omega$ Cen is the relic of a larger galaxy and was disrupted 
by the Galactic tidal field, Ferraro et al. (2002) suggested that the MMR stars were part 
of a small independent stellar system that has been accreted by the proto larger $\omega$ Cen. 
They propose this idea as an example of hierarchical merging on the subgalactic scale.

Alternatively, the observational results of cold kinematics and central concentration for more
metal-rich stars with respect to the metal-poor ones can lead to the speculation that the MMR
stars were formed from the chemically enriched gas which has selectively low angular momentum 
and was moved to the center of the cluster (Norris et al. 1997; Freeman 2002). This may be 
possible from the barlike asymmetry in the inner region which occurred in the process of the 
interactions between $\omega$ Cen and the Milky Way (Norris et al. 1997; Mayer et al. 2001a, 2001b). 
Close to the pericentric passage, $\omega$ Cen could develop a central bar, allowing part of 
the gas to funnel into the center of $\omega$ Cen and generating one burst of star formation. 
Stars in this central region could more efficiently lose their angular momentum because of 
the bar instability (Mayer et al. 2001a, 2001b). Probably the past ``tidally stirred" interaction 
between $\omega$ Cen and the Milky Way has disrupted the previously smooth and well-ordered 
distributions of the metal-poor populations in $\omega$ Cen and made it more efficient to trigger 
star formation of the MMR stars. If this is the case, the overall star formation of $\omega$ Cen 
must be an outside-in process. The initial star formation of the old MPP and MP populations 
occurred over the whole proto $\omega$ Cen far away from the young Milky Way, and susequent 
bursts of the young and more metal-rich populations were progressively more centrally concentrated, 
which may be dynamically connected to interactions with the Milky Way (Mayer et al. 2001a, 2001b; 
see also Strader et al. 2003 for the Fornax dwarf galaxy). If the MMR population was formed during 
the accretion to the Milky Way, the age of the MMR population may provide a hint of time taken 
for $\omega$ Cen to be accreted into the Galaxy. 

While the detailed mechanism of star formation in $\omega$ Cen is not currently clear,
all of above information strongly suggest that $\omega$~Cen is a relict or nucleus of 
a disrupted dwarf galaxy. The cases of $\omega$~Cen and the Sagittarius dwarf 
system therefore provide direct evidence for past and continuing accretion of protogalactic 
fragments, which suggest that similar accretion events may have continued throughout 
the history of Milky Way formation in the context of surviving ``building blocks'' of 
large galaxies (e.g., Searle \& Zinn 1978; Larson 1988; Zinn 1993).

\subsection{Other Globular Clusters with Metallicity Spread and Multiple Populations?}
Finally, it is of interest to note that $\omega$ Cen and M54, now proposed as possible nuclei 
or parts of disrupted dwarf galaxies, are respectively the most massive GC and second most 
massive GC in our Galaxy.  Thus it is interesting to examine other massive GCs for indications 
that they have similar histories, and of these NGC 6388 and NGC 6441 deserve detailed examination.  
They are the 3rd and 5th most massive GGCs, respectively, and they both have very peculiar bimodal 
HB distributions. In particular, they show color spreads among the RGB stars which appear to be larger 
than the effects expected from photometric errors and differential reddenings (Rich et al. 1997). 
Moreover, recent studies also have suggested a possible intrinsic spread in metallicity of these 
two clusters, although direct star by star metallicity determinations need to be made
(Piotto et al. 1997; Pritzl et al. 2001; Sweigart 2002).

Our population models for these unusual clusters show that the adoption of two distinct 
populations within the systems and very mild internal age-metallicity relations between the two 
populations can reproduce the observed features on the HB and the RGB (Ree et al. 2002; 
Yoon, Ree, and Lee 2003). The age and metallicity spreads, required to reproduce their CMD 
morphologies, are only about 1.2 Gyr and 0.15 dex for both clusters. This conclusion is also 
supported by the fact that the mean period of RR Lyrae variables in these clusters are too long 
for their high metallicities ([Fe/H] $\sim$ -0.5 - -0.6, Pritzl et al. 2000), which is understood 
in the two populations scenario if the RR Lyraes are highly evolved stars from the older and 
metal-poor BHB population.

It is worth noting that the most recent observations of the GC G1 in M31, which is
more than twice as massive as $\omega$ Cen, also shows
intrinsic metallicity dispersion (Meylan et al. 2001), implying a deep potential well enables 
massive clusters to retain enriched gas. It is suggestive therefore that there are additional 
GCs in the Milky Way and M31 that are the relicts of disrupted dwarf galaxies, 
as probably are $\omega$~Cen and M54.

\subsection{Classification of the Galactic Halo Globular Cluster System}
Our conclusions on the origin of $\omega$~Cen and other massive GCs, when combined with other
recent findings on the origin of the GGC system, suggests that the present day Galactic halo GCs 
can be subdivided into three different types:

(a) Clusters formed in a collapsing proto-Galaxy (e.g., Eggen, Lynden-Bell, \& Sandage 1962).
They are located mostly in the inner halo and are only genuine GGCs, in the classic sense.

(b) Clusters originally formed in satellite dwarf galaxies later accreted to the Milky Way.
They include GCs (Ter7, Ter8, Arp2, \& Pal12) belonging to the Sagittarius dwarf galaxy 
(Ibata, Irwin, \& Gilmore 1994; Ibata et al. 1997; Bellazzini, Ferraro, \& Ibata 2003 
and references therein), and young outer halo clusters with retrograde motion (Zinn 1993; 
van den Bergh 1993). Most recently, Yoon \& Lee (2002) also reported very compelling evidence 
that metal-poor ([Fe/H] $\la$ -2.0) Oosterhoff II clusters have the positional and orbital 
characteristics fully consistent with the hypothesis that they originated from a satellite 
galaxy such as the Magellanic Clouds.

(c) Nuclei (or relicts) of disrupted dwarf galaxies. $\omega$~Cen and M54, the nucleus of the
Sagittarius dwarf, are belonging to this type. Our population models suggest other massive GCs 
with bimodal HBs (e.g., NGC~6388 and NGC~6441) might also be included in this category.

\section{CONCLUSIONS}

In this paper, we present wide-field and high-precision $BV$ photometry for $\omega$ Cen,
which represents one of the most extensive photometric surveys yet obtained for this cluster.
We draw the following conclusions:

1. From our $Ca$ and Str\"omgren $by$ photometry, we removed foreground field populations
with $V$ $<$ 16.0 and [Fe/H] $\geq$ -0.6 dex in the CMD. The distinctly different distributions 
between $\omega$ Cen and the foreground field stars in the $hk$ vs. $b-y$ diagram, which is correlated 
with metallicity, has allowed us to discriminate the foreground field stars from 
the member stars of $\omega$ Cen and then to construct decontaminated CMDs.

2. The ``cleaned" CMDs of $\omega$ Cen show the presence of several RGBs. Notably, there 
exists a prominent feature of the MMR ([Fe/H] $\sim$ -0.5) RGB which is clearly separated and 
far redward from the main metal-poor RGBs. From the histograms of the RGB color distribution, 
we confirmed the existence of multiple stellar populations. The HB distribution is also consistent 
with the multimodal nature of the RGB. The evidence of discrete RGBs and corresponding HBs suggests 
that star formation in $\omega$ Cen occurred in successive bursts.

3. We identified, for the first time, a continuous and slanting RGB bump in the CMD of 
$\omega$ Cen, which is an unique feature among GGCs, and due to the metallicity difference 
between the stars. We found that multiple peaks of the RGB bump in the differential LF and 
the luminosities of these peaks depend on their metallicities, which is in good agreement 
with observational data among GGCs. All of these findings again allow us to confirm 
the existence of complex and multiple stellar populations with a wide spread in metallicity 
within $\omega$ Cen.

4. Our photometry has revealed a significant population of the BSSs in $\omega$ Cen. We found an 
intriguing group of stars, clearly separated from the main branch of the BSSs and 
lying above the cluster SGB, which may be stars evolved from the main BSS sequence. We also detected 
the BSS progeny candidates of $\omega$ Cen which are slightly brighter than the normal RHB stars and 
well separated from the AGB base. From the different shape of the observed BSS LFs for the inner 
and outer regions and comparison with the models, we found that there might be two different origins 
of the BSSs in $\omega$ Cen.

5. From our population models, we confirmed that four distinct populations with different 
metallicities can reproduce the observed multimodal nature of the RGB and the corresponding HB distributions. 
We determined the age-metallicity relation between the four distinct sub-populations 
in $\omega$ Cen by comparing the observed HB distribution with our population models. We suggest 
that the MMR ([Fe/H] $\sim$ -0.5) population is a few billion years younger ($\Delta t$ $\sim$ 4 Gyr) 
than the MMP ([Fe/H] $\sim$ -1.8) population. A consistent picture of the evolved stars 
in $\omega$ Cen is given by the presence of (1) a metal-poor population of [Fe/H] $\sim$ -1.8 - -1.3 
and age $\sim$ 12 Gyr with exclusively BHB and blue side of the RGB (MMP and MP sub-populations), 
(2) a metal-rich population of [Fe/H] $\sim$ -1.0 and age $\sim$ 11 Gyr with the RHB and red side 
of the RGB (MR sub-population), and (3) the most metal-rich population of [Fe/H] $\sim$ -0.5 and 
age $\sim$ 8 Gyr required to explain the distinctly separated RGB sequence and its RHB clump 
visible in the more metal-poor RGB bump region (MMR sub-population). From the discovery of 
several distinct populations and the internal age-metallicity relation found in our work and 
others (Lee et al. 1999; Pancino et al. 2000; Hilker \& Richtler 2000; Hughes \& Wallerstein 2000), 
we suggest $\omega$ Cen was once part of a more massive system that merged with the Milky Way, 
as the Sagittarius dwarf galaxy is in the process of doing now.

New observational data are  required to confirm our conclusions presented above.  In particular,
detection of any  RHB clump stars residing amongst 
more metal-poor RGB stars would be possible via a spectroscopic and/or spectrophotometric 
(e.g., $uvby$ Str\"omgren photometry) survey for stars around the RHB clump region in the CMD. 
From derived metallicity and gravity values, the RHB clump stars could be
directly discriminated from the more metal-poor RGB stars. This would clarify the 
age-metallicity relation suggested by our models, which is based on the location of 
the RHB clump in the CMD.

\acknowledgments
We thank F. Ferraro for providing the RGB fiducial lines of Galactic globular clusters. 
We also thank N. Suntzeff for providing an electronic dataset of [Fe/H] values of giants 
in $\omega$ Cen. We would like to thank the anonymous referee for a careful review and
many helpful suggestions. Support for this work was provided by the Creative Research 
Initiatives Program of Korean Ministry of Science \& Technology.

\clearpage

\begin{figure}
 \caption{The observed cluster region of $\omega$ Cen.}
\end{figure}

\begin{figure}
\centerline{\epsfysize=7in%
\epsffile{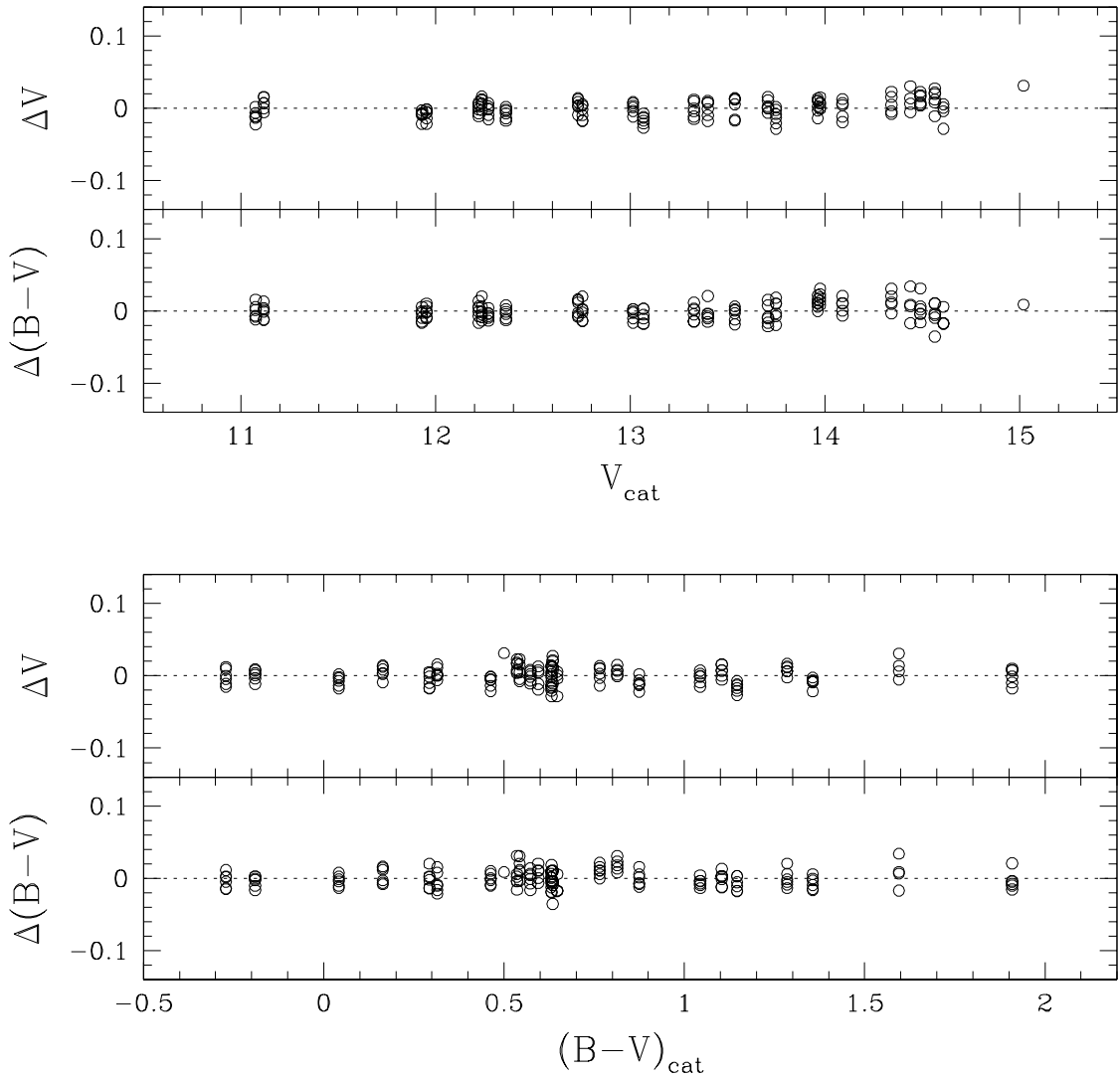}}
 \caption{Magnitude and color residuals for the comparison of Landolt (1992)
standards and measured values in this study, in the sense of our values minus 
those of Landolt (1992).}
\end{figure}

\begin{figure}
\centerline{\epsfysize=7in%
\epsffile{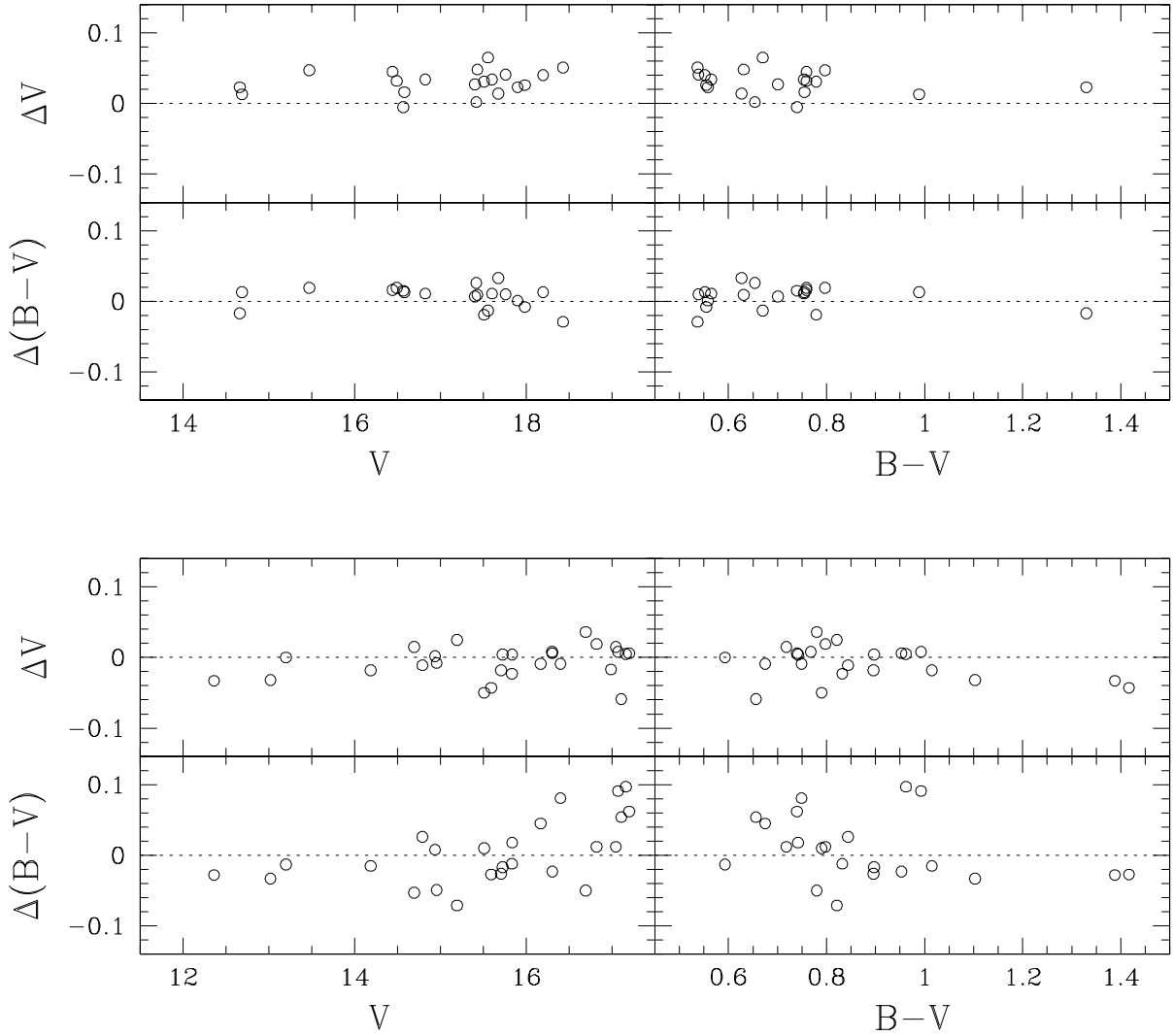}}
 \caption{Comparison of our photometry with those of ($upper$) Walker (1994) and 
($lower$) Alcaino \& Liller (1987). The differences are in the sense of our photometry
minus others.}
\end{figure}

\begin{figure}
 \caption{$V$, $B-V$ CMDs for each of the nine fields of $\omega$ Cen. 
The field F5 corresponds to the one placed on the cluster center. 
Only stars detected on twenty or more frames in each filter are shown. 
No other restrictions have been imposed.}
\end{figure}

\begin{figure}
 \caption{$V$, $B-V$ CMD for all stars in Fig. 4.}
\end{figure}

\begin{figure}
 \caption{
($upper$) $hk$ vs. $b - y$ diagram of stars in the program field with $V$ $<$ 16 mag. 
Note that the Galactic field star candidates (closed circles) are discriminated 
from the evolved stars (dots) of $\omega$ Cen in this diagram. This is due to the substantial 
difference between their mean metallicities. The upper envelope of these field star candidates
cover empirical isometallicity lines for the main-sequence dwarfs with [Fe/H] = -0.5 and 0.0 
(solid lines, Anthony-Twarog et al. 2000). The RR Lyrae stars are represented by open circles
whose data was obtained from Rey et al. (2000). The BHB, RHB, and RGB stars are also indicated.
($lower$) $V$, $B-V$ CMD for all observed stars matched with $Ca$ \& Str\"omgren $by$ data
and selected with photometric errors less than 0.05 mag in $B$ and $V$. The selected field star 
candidates (closed circles) show a prominent sequence, as expected, running through 
the middle of the $\omega$ Cen CMD.
}
\end{figure}

\begin{figure}
 \caption{Cleaned CMD for stars matched with $Ca$ \& Str\"omgren $by$ data
and selected with photometric errors less than 0.05 mag in $B$ and $V$. 
A subset of the Galactic field stars have been subtracted using 
the $hk$ vs. $b-y$ diagram.}
\end{figure}

\begin{figure}
 \caption{(a) Zoomed CMD of the RGB region. In order to construct histograms of color distribution,
the RGB stars are selected in a relatively narrow magnitude ranges shown by dashed lines.
The arrow indicates the MMR RGB, well separated from the bulk of bluer metal-poor stars.
(b) and (c) Color distributions of the RGB stars for two magnitude ranges. The number of four main 
components (1: MMP RGB, 2: MP RGB, 3: MR RGB, and 4: MMR RGB) and their [Fe/H] abundances are 
indicated.}
\end{figure}

\clearpage
\begin{figure}
 \caption{Four sub-populations of the RGB stars in the CMD. The RGB sub-populations 
(i.e., MMP, MP, MR, and MMR RGBs) are divided by five RGB sequences (dashed lines) whose 
metallicities and shapes are determined by a semi-empirical method (see text). The common stars 
observed by Suntzeff \& Kraft (1996) and us are marked using different symbols for different 
metallicity groups.}
\end{figure}

\begin{figure}
 \caption{
($upper$) CMD for $\omega$ Cen. We overplot polynomial fitted lines that differentiate
seven RGB groups with different metallicities. From [($B-V$)$_{o}$, $M_{V}$] $\sim$ (0.8, 14.3) 
to (0.9, 15.5), there is an evident inclined sequence of the RGB bumps which is mainly 
due to the continuous spread of metallicity. ($lower$) The differential LFs of the seven 
RGB groups with different metallicities. The RGB bump for each RGB group is clearly seen. 
The arrow indicates our adopted magnitude of each RGB bump.}
\end{figure}

\begin{figure}
 \caption{
(a) The parameter $\Delta V_{ZAHB}^{bump}$ as a function of metallicity in Zinn \& West (1984) 
scale for seven groups of the $\omega$ Cen RGB bump (closed circles). The data for 42 GGCs compiled 
by Ferraro et al. (1999) are plotted as open circles. The dotted line is the regression 
through these GGCs data. (b) Same as (a) but for absolute magnitude [$M_{V}(bump)$] of the RGB bump 
as a function of metallicity. Note that the behavior of the luminosities of the $\omega$ Cen RGB bumps 
as a function of metallicity is in good agreement with that found for GGCs.
}
\end{figure}

\begin{figure}
 \caption{
CMD for the HB and the BS stars of $\omega$ Cen. The RR Lyrae stars from our photometry are plotted 
as open circles. Subtraction of the field stars has allowed us to obtain accurate distribution of 
the RHB stars. The RHB and the BSS progeny region is denoted by a dashed line and solid line box, respectively. The distribution of the BSSs (closed circles) is compared with the ZAMS (solid line) 
and sequences with various ages (2, 3, 4, 5, and 6 Gyr, long dashed lines; 12 Gyr, gray line) from 
the Y$^{2}$ isochrones for the Y = 0.23 and Z = 0.0006, which is consistent with the chemical abundance of the MMP population of $\omega$ Cen. The YSSs are shown by crosses.
}
\end{figure}

\begin{figure}
\centerline{\epsfysize=7in%
\epsffile{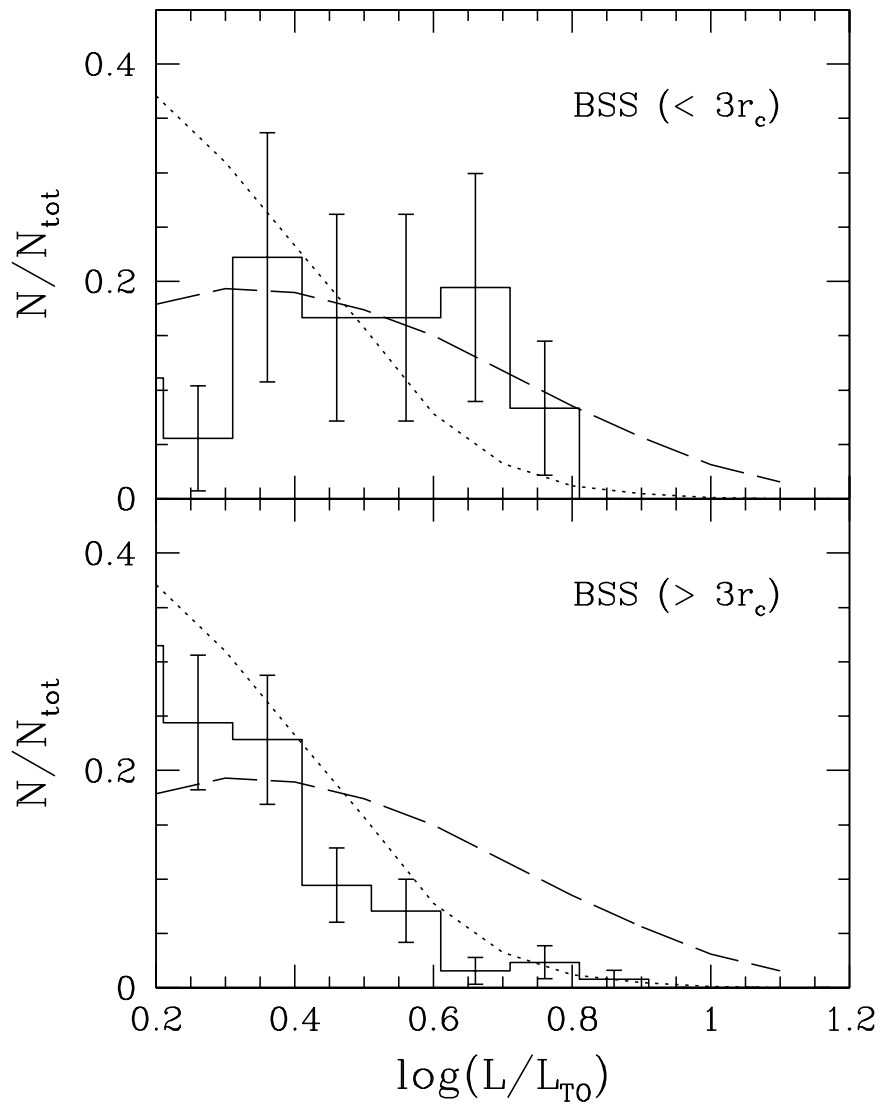}}
 \caption{
The BSS LFs for the inner ($upper$) and outer region ($lower$). These LFs are compared with two
theoretical models for the BSS formation: stellar collisions (dashed line) and 
primordial binary mergers (dotted line), obtained from Bailyn \& Pinnsonault (1995).
The two LFs corresponding to the inner and outer regions of $\omega$ Cen appear to be different 
from each other.
}
\end{figure}

\clearpage
\begin{figure}
\centerline{\epsfysize=7in%
\epsffile{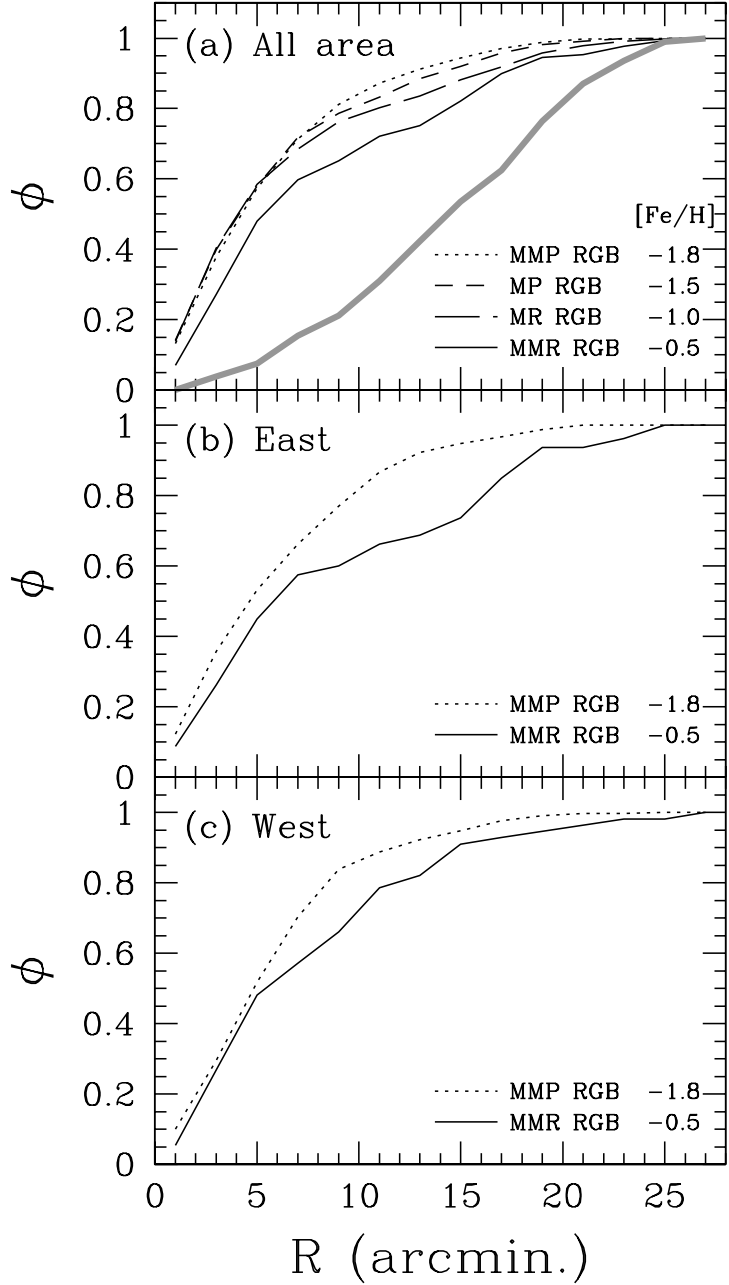}}
 \caption{
(a) The cumulative radial distributions of four RGB sub-populations with different metallicities.
The thick gray line represents a distribution of the probable foreground field stars extracted from 
our $Ca$ \& Str\"omgren $by$ photmetry. The radial distribution of the MMR RGB appears to be different from the more metal-poor ones. The cumulative radial distributions of the MMP and the MMR RGB stars in (b) the east and (c) the west field. The MMR RGB stars in the east field show a more prominent bimodal distribution compared to that in the west field. 
}
\end{figure}

\begin{figure}
\centerline{\epsfysize=7in%
\epsffile{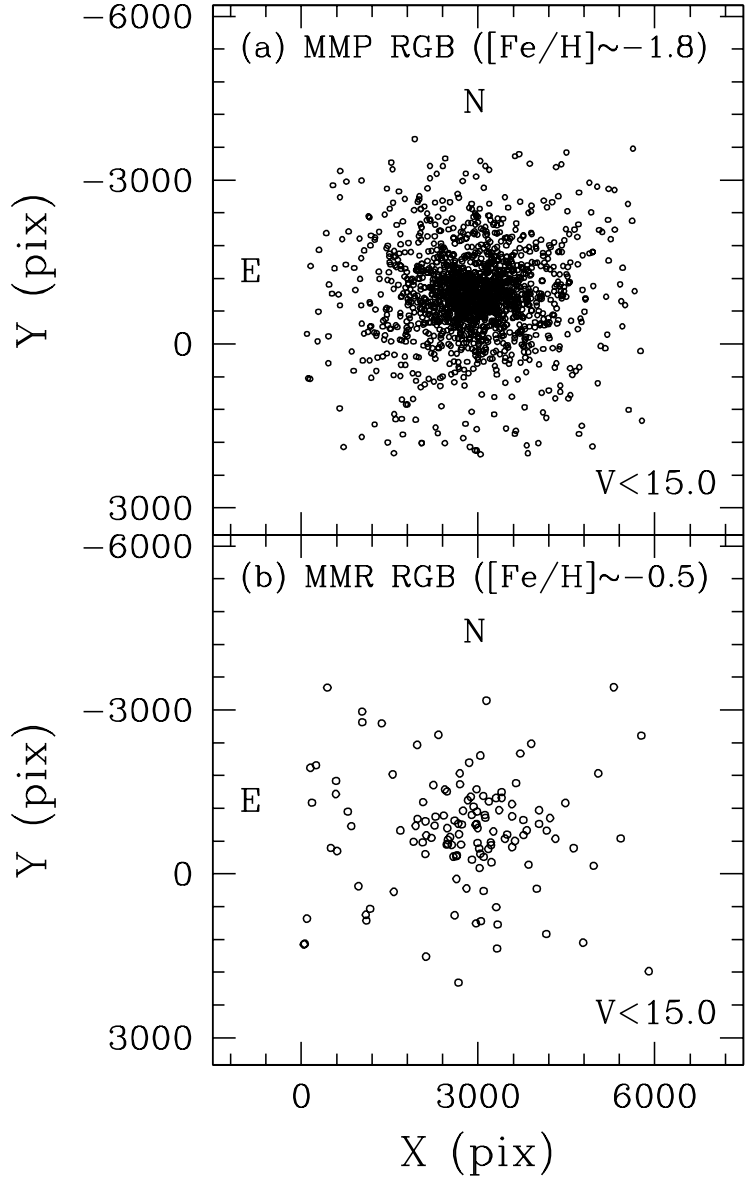}}
 \caption{
Spatial distributions of (a) the MMP RGB and (b) the MMR RGB stars with $V$ $<$ 15.0. 
There are more MMR RGB stars at higher radii to the east.
}
\end{figure}

\begin{figure}
\centerline{\epsfysize=7in%
\epsffile{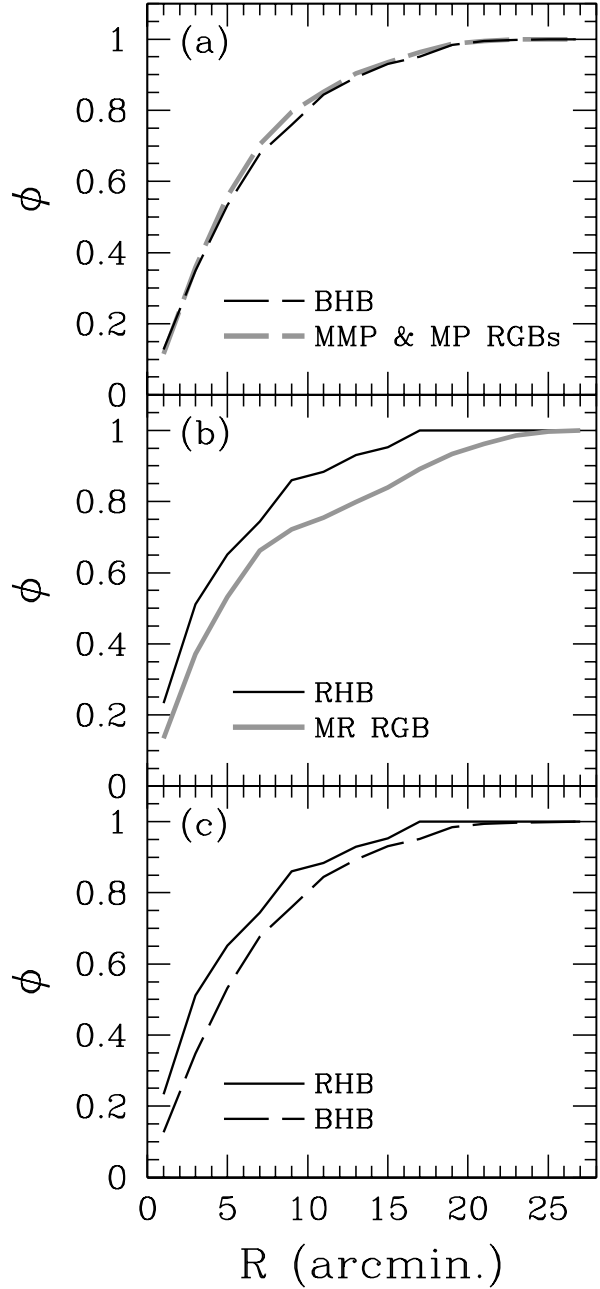}}
 \caption{
The cumulative radial distributions of the BHB and the RHB with respect to those of the corresponding 
RGBs with different metallicities of [Fe/H] $\sim$ -2.0 - -1.3 (i.e., MMP and MP RGBs) and
[Fe/H] $\sim$ -1.0 (i.e., MR RGB). (a) While the BHB shows a similar distribution with that of 
the MPP and the MP RGB populations, (b) the radial distribution of the RHB is somewhat different from 
that of the corresponding MR RGB. (c) The RHB shows a more central concentration than the BHB.
}
\end{figure}

\begin{figure}
\centerline{\epsfysize=7in%
\epsffile{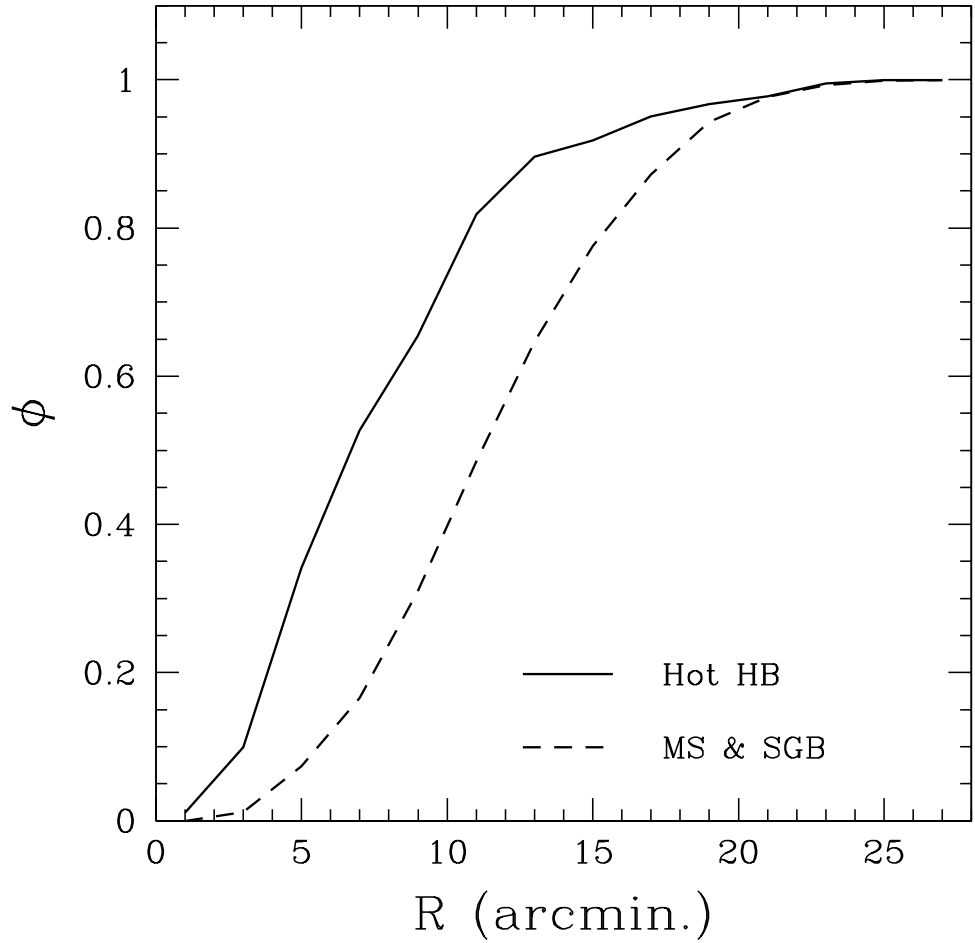}}
 \caption{
The cumulative radial distribution of stars located in the hot and faint HB (16.0$<V<$18.4, -0.40$<B-V<$0.15). We also plot a distribution of the MS and the SGB stars in the same magnitude 
range of the hot HB stars. The hot HB stars are apparently more concentrated toward the cluster center 
than are the MS and the SGB stars. 
}
\end{figure}

\begin{figure}
\centerline{\epsfysize=7in%
\epsffile{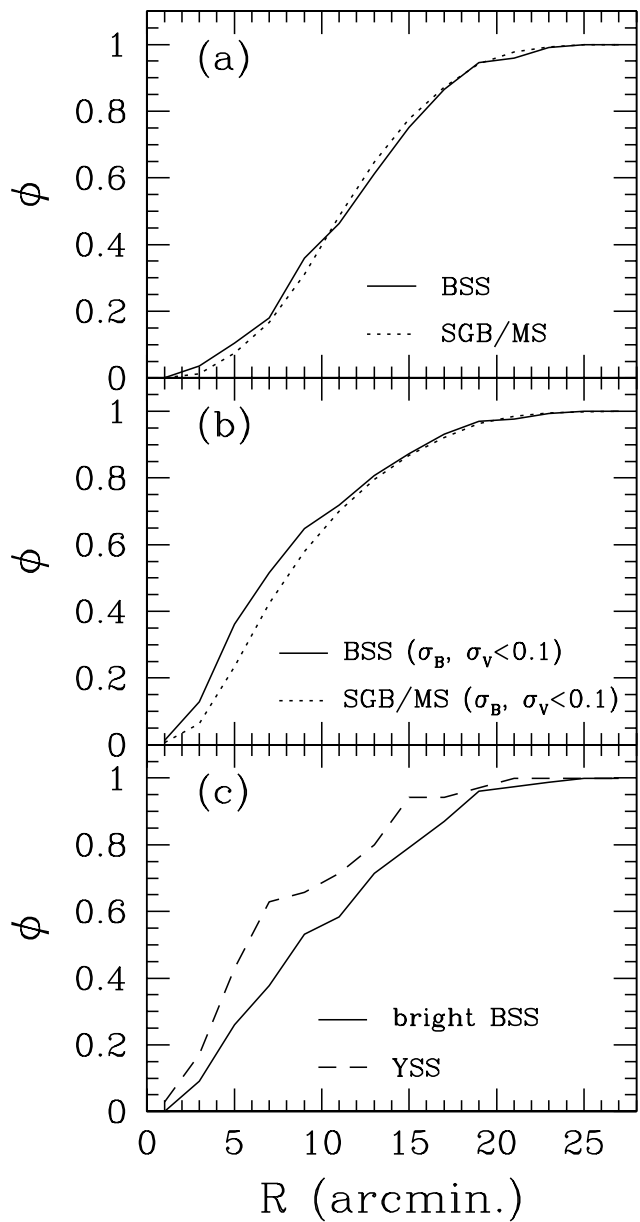}}
 \caption{
(a) The cumulative radial distribution of the BSSs compared with the SGB/MS stars in a similar
magnitude range. (b) Same as (a), but for the BSSs and the SGB/MS stars with a frame-to-frame error 
in the $B$ and $V$ magnitudes of less than 0.1 mag. (c) Same as (a), but for the bright 
(16.0 $<$ $V$ $<$ 17.2) BSSs and the YSSs.
}
\end{figure}

\clearpage
\begin{figure}
 \caption{
The HB population models for $\omega$ Cen (MMP, MP, MR, and MMR) plotted over the observed 
CMD. We also overplot $Y^{2}$ isochrones representing four distinct populations.
The instability strip is schematically represented by solid lines. Four populations 
with different metallicities and ages can reproduce the observed discrete nature of the RGB 
and the HB morphology (see text). The RHB clump associated with the MMR RGB is located in the region overlapping with the RGB bumps of the more metal poor populations.
}
\end{figure}

\begin{figure}
 \caption{
The models illustrate the estimation of age difference between the BHB associated with
the MMP ($Z$ = 0.0006) component and the RHB clump associated with the MMR ($Z$ = 0.006) 
population assuming the fixed age (12 Gyr) for the MMP population. Note the variation 
in mean colors of the RHB clumps as well as in the MSTO luminosities of the MMR population.
}
\end{figure}

\begin{figure}
 \caption{
($upper$) A CMD that highlights the RGB and the RGB bump region. Three RGB groups with 
different metallicities are shown. The RGB bump is shaped as an inclined sequence. 
The predicted location from our HB models of the RHB clump with [Fe/H] $\sim$ -0.5, 
which is younger ($\Delta t$ $\sim$ 4 Gyr) than the MMP population, is indicated. 
($lower$) Normalized LFs for the metal-poor (dashed line) and the intermediate metallicity 
(solid line) RGBs. Peaks of the two LFs correspond to the RGB bumps. Note that the peak of 
the intermediate metallicity RGB bump appears to be higher than that of the metal-poor one, 
indicating that the RHB clump stars might enhance the  intermediate metallicity RGB bump 
feature due to their similar locations. An arrow indicates the predicted location of 
the RHB clump with [Fe/H] $\sim$ -0.5 from our models, which is younger ($\Delta t$ $\sim$ 4 Gyr) 
than the MMP population.
}
\end{figure}

\clearpage

\begin{deluxetable}{cccl}

\setcounter {table}{0}
\tablewidth{0pt}
\tablecaption{Position of $\omega$ Cen fields}
\tablehead{
\colhead{Field}       &      \colhead {RA (2000)}  &
\colhead {DEC (2000)}   &    \colhead {Comment}}
\scriptsize
\startdata
     F1   &  13:28:04.60  &   -47:41:32.10  &  SE  \nl
     F2   &  13:28:04.90  &   -47:28:30.30  &  E   \nl
     F3   &  13:28:05.20  &   -47:15:30.20  &  NE  \nl
     F4   &  13:26:48.60  &   -47:15:30.20  &  N   \nl
     F5   &  13:26:47.90  &   -47:28:38.70  &  Cluster center \nl
     F6   &  13:26:48.10  &   -47:41:33.70  &  S   \nl
     F7   &  13:25:31.00  &   -47:41:35.60  &  SW  \nl
     F8   &  13:25:30.50  &   -47:28:32.90  &  W   \nl
     F9   &  13:25:30.00  &   -47:15:35.30  &  NW  \nl   
     
\enddata
\end{deluxetable}

\clearpage

\begin{deluxetable}{lrrcccc}

\setcounter {table}{1}
\tablewidth{0pt}
\tablecaption{CCD Photometry for member stars of $\omega$ Cen{*}}
\tablehead{
\colhead{ID}  & \colhead{$x$}  &   \colhead{$y$}  &   \colhead{$V$}   &   \colhead{$\sigma_{V}$}
&    \colhead{$B-V$}   &    \colhead{$\sigma_{(B-V)}$}}
\startdata
        1    &   3702.16    &   1874.11    &    9.846    &    0.009    &    1.034    &    0.010 \nl
        2    &   4769.50    &    497.98    &   10.009    &    0.006    &    1.286    &    0.007 \nl
        3    &   1754.23    &   1229.86    &   10.041    &    0.010    &    1.348    &    0.010 \nl
        4    &   4671.58    &    -63.43    &   10.050    &    0.004    &    1.141    &    0.004 \nl
        5    &    136.88    &    409.37    &   10.288    &    0.005    &    1.683    &    0.005 \nl
\enddata
\tablenotetext{*} {The complete version of this table is in the electronic edition of the Journal.
The printed edition contains only a sample.}
\end{deluxetable}

\begin{deluxetable}{lrrcccc}

\setcounter {table}{2}
\tablewidth{0pt}
\tablecaption{CCD Photometry for field star candidates with $V$ $<$ 16.0{*}}
\tablehead{
\colhead{ID}  & \colhead{$x$}  &   \colhead{$y$}  &   \colhead{$V$}   &   \colhead{$\sigma_{V}$}
&    \colhead{$B-V$}   &    \colhead{$\sigma_{(B-V)}$}}
\startdata
        1    &   1041.68    &  -1702.05    &    9.458    &    0.008    &    0.460    &    0.013 \nl
        2    &   4973.86    &  -1018.14    &   10.406    &    0.008    &    0.295    &    0.009 \nl
        3    &   2198.69    &    319.31    &   10.499    &    0.004    &    0.573    &    0.005 \nl
        4    &   1825.23    &  -3152.36    &   10.514    &    0.008    &    0.478    &    0.008 \nl
        5    &    270.93    &   1119.84    &   10.770    &    0.003    &    0.531    &    0.004 \nl
\enddata
\tablenotetext{*} {The complete version of this table is in the electronic edition of the Journal.
The printed edition contains only a sample.}
\end{deluxetable}

\clearpage
\begin{deluxetable}{ccrc}

\setcounter {table}{3}
\tablewidth{0pt}
\tablecaption{RGB Bump Parameters of $\omega$ Cen}
\tablehead{
\colhead{RGB bump}       &      \colhead{$V_{bump}$}  &
\colhead{$\Delta V_{ZAHB}^{bump}$}        & \colhead{[Fe/H]$_{ZW}$}}
\scriptsize
\startdata
     1     &  14.28 $\pm$ 0.05   &   -0.44 $\pm$ 0.11  &  -1.92 $\pm$ 0.14         \nl
     2     &  14.35 $\pm$ 0.05   &   -0.36 $\pm$ 0.11  &  -1.74 $\pm$ 0.11         \nl
     3     &  14.50 $\pm$ 0.05   &   -0.20 $\pm$ 0.11  &  -1.56 $\pm$ 0.14         \nl
     4     &  14.55 $\pm$ 0.05   &   -0.15 $\pm$ 0.11  &  -1.37 $\pm$ 0.13         \nl
     5     &  14.73 $\pm$ 0.05   &    0.02 $\pm$ 0.11  &  -1.19 $\pm$ 0.13         \nl
     6     &  14.93 $\pm$ 0.05   &    0.21 $\pm$ 0.11  &  -0.94 $\pm$ 0.21         \nl
     7     &  15.45 $\pm$ 0.05   &    0.55 $\pm$ 0.09  &  -0.51 $\pm$ 0.29         \nl
\enddata
\end{deluxetable}

\clearpage

\begin{deluxetable}{clcllll}

\setcounter {table}{4}
\tablewidth{0pt}

\tablecaption{Age-Metallicity Relation of $\omega$ Cen}
\tablehead{
\colhead{} &  \colhead{Population}    &    \colhead{[Fe/H]}  &
\colhead{Z}     &  \colhead{Age (Gyr)}   & \colhead{RR Lyraes} &   \colhead{Remark}}
\scriptsize
\startdata
1 & MMP RGB and BHB  &  -1.8  &  0.0006  &  12   &  many  &  age from $Y^{2}$ isochrones  \nl
2 & MP RGB and BHB   &  -1.5  &  0.001   &  12   &  many  &  $\Delta t$ from HB models    \nl
3 & MR RGB and RHB   &  -1.0  &  0.002   &  10.8 &  few   &  $\Delta t$ from HB models    \nl
4 & MMR RGB and RHB clump &  -0.5 & 0.006 &  8   &  none  &  $\Delta t$ from HB models and RGB LF \nl
\enddata
\end{deluxetable}

\end{document}